%% file: main.tex
\documentclass[10pt,twocolumn,letterpaper]{article}

\usepackage[pagenumbers]{cvpr}      

\input{preamble}

\definecolor{cvprblue}{rgb}{0.21,0.49,0.74}
\usepackage[pagebackref,breaklinks,colorlinks,citecolor=cvprblue]{hyperref}
\usepackage{amsmath}
\usepackage{amssymb}
\usepackage{amsmath}
\usepackage{amssymb}
\usepackage{booktabs}
\usepackage{multirow}
\usepackage{bbding}
\usepackage{textcomp}
\usepackage{tablefootnote}
\usepackage[accsupp]{axessibility}
\usepackage{hyperref}
\usepackage{graphicx}
\hypersetup{pagebackref,breaklinks,colorlinks}
\newcommand{\ieno}{\textit{i.e.}}
\newcommand{\egno}{\textit{e.g.}}
\def\paperID{5577} 
\def\confName{CVPR}
\def\confYear{2024}
\newcommand{\ours}{\textit{KSVQE}}
\newcommand{\tcr}{\textcolor{black}}
\newcommand{\tcb}{\textcolor{black}}
\newcommand{\tct}{\textcolor{black}}

\title{KVQ: Kwai Video Quality Assessment for Short-form Videos}
\author{Yiting Lu\textsuperscript{\rm 1}\footnotemark[1], Xin Li\textsuperscript{\rm 1}\footnotemark[1], Yajing Pei\textsuperscript{\rm 1,2}\footnotemark[1], Kun Yuan\textsuperscript{\rm 2}\footnotemark[2], \\
Qizhi Xie\textsuperscript{\rm 2,3}, Yunpeng Qu\textsuperscript{\rm 2,3}, Ming Sun\textsuperscript{\rm 2}, Chao Zhou\textsuperscript{\rm 2}, Zhibo Chen\textsuperscript{\rm 1}\footnotemark[2]\\
\textsuperscript{\rm 1}University of Science and Technology of China,  \textsuperscript{\rm 2}Kuaishou Technology,
\textsuperscript{\rm 3}Tsinghua University \\
{\tt \small \{luyt31415,lixin666,peiyj\}@mail.ustc.edu.cn,} 
{\tt \small chenzhibo@ustc.edu.cn} \\
{\tt\small \{yuankun03,xieqizhi,quyunpeng,sunming03,zhouchao\}@kuaishou.com}}

\begin{document}
\twocolumn[{
\renewcommand\twocolumn[1][]{#1}%
\maketitle
\vspace{-6mm}
\begin{center}
\captionsetup{type=figure}
    \setlength{\abovecaptionskip}{-0.001cm} 
    \includegraphics[width=1\textwidth]{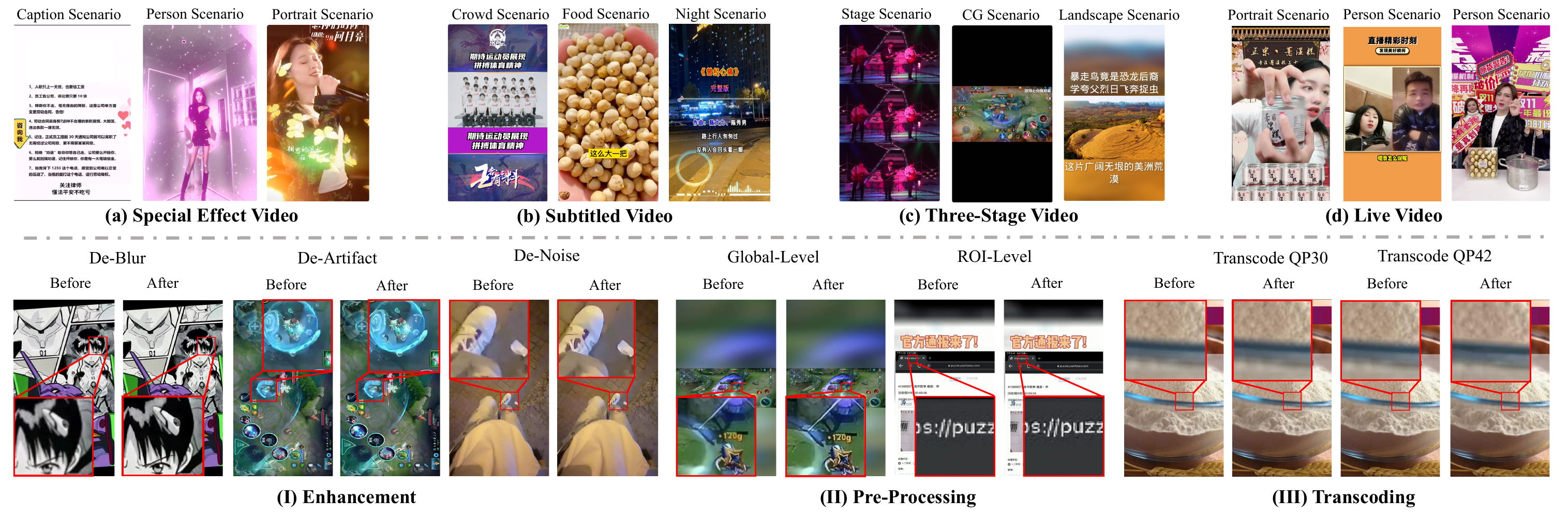}
    \captionof{figure}{The two primary challenges of short-form videos: \textit{the kaleidoscope content} with various creation modes (top) and \textit{complicated distortion} arising from sophisticated video processing workflows (bottom). Regions with distortions are indicated by red boxes.
    }
    \label{fig:first}
\end{center}
}] 

\maketitle

\input{sec/0_abstract}
\vspace{-3mm}
\input{sec/1_intro1}
\input{sec/2_Related_work}
\input{sec/3_database}

\input{sec/4_method}

\input{sec/5_experiment}
\input{sec/6_conlusion}

\clearpage

{
    \small
    \bibliographystyle{ieeenat_fullname}
    \bibliography{main}
}

\input{sec/X_suppl}

\end{document}

%% file: preamble.tex
\usepackage[dvipsnames]{xcolor}


%% file: sec/0_abstract.tex
\renewcommand{\thefootnote}{\fnsymbol{footnote}}
\footnotetext[1]{Equal contribution}
\footnotetext[2]{Corresponding authors}

\begin{abstract}
Short-form UGC video platforms, like Kwai and TikTok, have been an emerging and irreplaceable mainstream media form, thriving on user-friendly engagement, and kaleidoscope creation, etc. However, the advancing content-generation modes, \egno, special effects, and sophisticated processing workflows, \egno, de-artifacts, have introduced significant challenges to recent UGC video quality assessment: (i) the ambiguous contents hinder the identification of quality-determined regions. (ii) the diverse and complicated hybrid distortions are hard to distinguish. 
To tackle the above challenges and assist in the development of short-form videos, we establish the first large-scale \textbf{K}aleidoscope short \textbf{V}ideo database for \textbf{Q}uality assessment, termed KVQ, which comprises 600 user-uploaded short videos and 3600 processed videos through the diverse practical processing workflows, including pre-processing, transcoding, and enhancement. Among them, the absolute quality score of each video and partial ranking score among indistinguishable samples are provided by a team of professional researchers specializing in image processing. Based on this database, we propose the first short-form video quality evaluator,  \ieno, \textit{KSVQE}, which enables the quality evaluator to identify the quality-determined semantics with the content understanding of large vision language models (\ieno, CLIP) and distinguish the distortions with the distortion understanding module. 
Experimental results have shown the effectiveness of \ours~on our KVQ database and popular VQA databases. The project can be found at  ~\url{https://lixinustc.github.io/projects/KVQ/}.
\end{abstract}

%% file: sec/1_intro1.tex
\vspace{-3mm}
\section{Introduction}

\tcr{Recent years have witnessed the significant advancement of short-form UGC video platforms, where billions of users have actively engaged in uploading and sharing their user-generated content (UGC) videos that encompass personal life, professional skills, and education, etc. Different from traditional video platforms, such as YouTube, short-form video platform aims to simplify content creation for users and enhance the accessibility and conciseness of video content for viewers by limiting the video length, which achieves great success since their mobile-friendly broadcasting, user-friendly engagement, kaleidoscope content creation, and snackable content. Despite that, the variable and uncertain subjective quality caused by non-professional shooting~\cite{Internet,SPAQ} or bitrate constrain~\cite{DBLP:conf/icip/PavezPXOA22,compression,RLVC} urgently entails the development of the video quality assessment (VQA) tailored for the short-form UGC (S-UGC) videos. }

\tcb{Recently, most existing databases~\cite{UGC-VQA, youtubeUGC,LSVQ,KoNViD-1k, maxwell} and associated studies~\cite{GSTVQA,VQT,simpleVQA,Dover,pamifastvqa,tpqi,LSCT} for the UGC video quality assessment are contributed for the in-the-wild UGC videos from general media platforms (\egno, Youtube). And these excellent databases can be divided into two main streams. One of the streams~\cite{youtubeUGC, LSVQ,KoNViD-1k} merely focused on the quality of UGC videos acquired from traditional stream media clients. Another line of these UGC databases~\cite{UGC-VIDEO,MDVQA} delved into the impact of compression on UGC videos.}
\tcr{In contrast, there are two primary challenges for the quality assessment of S-UGC videos that prevent the application of existing UGC methods: (i) the presence of various special creation/generation modes, \egno, special effects (Please see Fig.~\ref{fig:first}) and kaleidoscope contents, including portrait, landscape, food, etc, which confuses and impede the VQA models to accurately identify the quality-determined region/contents. (ii) sophisticated processing flow, \egno, transcoding and enhancement, along with intricate distortions existing in user-uploaded videos, which presents significant difficulties for the VQA model in distinguishing and determining the video quality. }

\tcr{To further improve the quality assessment of S-UGC videos, we establish the first large-scale kaleidoscope short-form video database named KVQ. In particular, 4200 S-UGC videos are collected to cover the primary creation modes (\egno, special effect and three-stage form) and content scenarios (\egno, food, stage, night, and so on) in the popular short-from UGC video platform, which is composed of 600 user-uploaded S-UGC videos and 3600 processed S-UGC videos via several practical video processing workflows~\cite{enhancement1,enhancement2,transcode,transcode1} (\egno, pre-processing, enhancement, transcoding). \textit{Notably, the selection of content and processing strategies are determined by practical statistics in the popular S-UGC platform, which is significant for the development and measurement of S-UGC VQA.} To provide accurate annotation for KVQ, a team of professional researchers specializing in image processing is responsible for the quality labeling of each S-UGC video with the range of [1-5] and the interval of 0.5. Despite that, there are still some videos with similar subjective quality, which makes it hard to distinguish which is better. To empower our KVQ with more fine-grained quality estimation capability, we select 500 indistinguishable S-UGC video pairs and provide their ranked annotations, which are not considered by existing UGC datasets.}

\tcr{Based on our KVQ benchmark, we introduce the first Kaleidoscope Short-form UGC Video Quality Evaluator (KSVQE). In particular, to identify the quality-determined regions and mitigate the impacts of quality-unrelated content, it is necessary to enhance the content understanding capability of our KSVQE. Considering the powerful fine-grained semantic understanding capability of pre-trained large vision-language model, CLIP~\cite{CLIP}, we propose the quality-aware region selection module (QRS) and content-adaptive modulation (CaM) for KSVQE. In QRS, the learnable quality adapter is introduced to adapt the fine-grained semantics from pre-trained CLIP as the guidance to identify the quality-determined regions and keep it, while dropping the quality-unrelated contents. The CaM is introduced to enable our KSVQE to perceive the content semantics for each region, since the subjective quality is also associated with different contents. 
To address the indistinguishability of distortions in S-UGC videos caused by video shooting and sophisticated processing workflows, we enhance the distortion understanding and adaptation capability of our KSVQE, by incorporating the distortion prior captured with the distortion-aware model CONTRIQUE~\cite{CONTRIQUE}. Here, the CONTRIQUE is efficiently fine-tuned toward the distortion distribution of our KVQ database with a distortion adapter under the contrastive loss function. With the above innovations, our KSVQE achieves state-of-the-art performance on our proposed KVQ dataset, which excessively outperforms the current best method Dover (\textit{retrained with our KVQ}) by 0.032 on PLCC and 0.034 on SROCC. Moreover, our proposed KSVQE owns great applicability for the commonly-used UGC-VQA datasets.} 
\tcr{The contributions of this paper are summarized below:
\begin{itemize}
    \item We built the first large-scale kaleidoscope short-form video database, termed KVQ, which is composed of 4200 user-uploaded or processed short-form videos collected from the popular short-form UGC video platform. 
    The reliable absolute quality label and partial ranked label for indistinguishable samples are annotated by a group of professional researchers specializing in image processing. 
    \item We propose the first kaleidoscope short-form video quality evaluator, termed KSVQE, to solve two primary challenges in KVQ: (i) unidentified quality-determined region/content caused by various creation/generation modes and kaleidoscope content scenarios. (ii) indistinguishable distortions caused by sophisticated processing flows and unprofessional video shooting. 
    \item To enable the content understanding capability of KSVQE, we propose the quality-aware region selection module (QRS) and content-adaptive modulation (CaM) based on the pre-trained large vision-language model, CLIP. Apart from that, we enhance the distortion understanding of KSVQE by designing the distortion-aware modulation (DaM) via a pre-trained distortion extractor. 
    \item The thorough analysis of our KVQ is provided and extensive experiments on our proposed KVQ and the commonly-used UGC VQA datasets have shown the effectiveness and applicability of our proposed KSVQE. 
\end{itemize}}

%% file: sec/2_Related_work.tex
\vspace{-4mm}
\section{Related Work}
\label{sec:Related_Work}
\begin{figure*}[htp]
    \centering
    \includegraphics[width=1\textwidth]{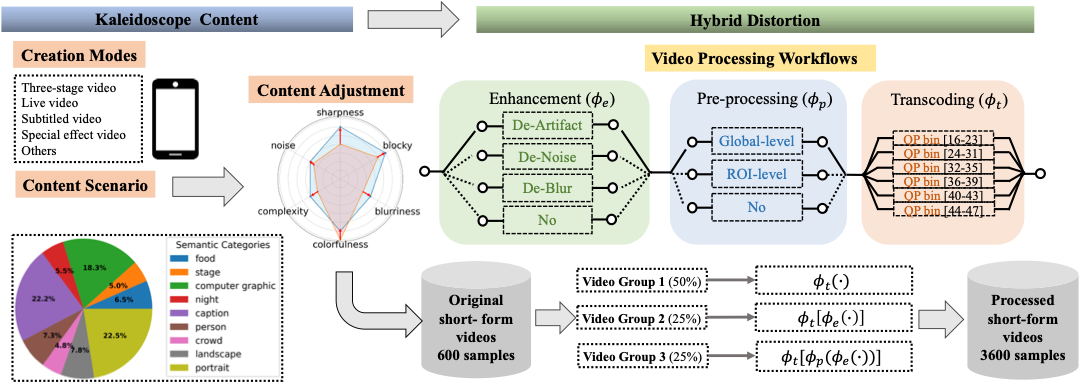}
    \caption{The overview for establishing the KVQ dataset involves several key steps. Initially, we collect the original short-form videos to cover the primary creation modes and content scenarios. Subsequently, we make fine-grained video content adjustments based on the 6 video features. Finally, sophisticated video processing workflows are applied to incorporate various hybrid distortions.} 
    \label{fig:framework}
\end{figure*}
\subsection{UGC-VQA databases}
In recent years, to develop more realistic and challenging video quality assessment (VQA) for user-generated content (UGC), many UGC databases~\cite{CVD2014,UGC-VQA,LIVE-Qualcomm,LIVE-VQC,LSVQ,MDVQA,maxwell}, have been established collecting videos with authentic distortions. These databases can be categorized into two types based on their collection scope. 
The first category~\cite{KoNViD-1k,youtubeUGC,LSVQ}contains UGC databases collected from the real-world media platform. Notably, LSVQ~\cite{LSVQ} includes a substantial 39,076 videos. The Second category~\cite{UGC-VIDEO,MDVQA} involves UGC databases with simulated distortions approximating realistic online video platforms, containing both originally distorted and post-compressed videos. Our proposed KVQ database, gathered from a short video platform, is similar to the Second category but has two key differences. Firstly, KVQ focuses extensively on short-form videos with various creation modes and kaleidoscope content. Secondly, KVQ underwent sophisticated video processing workflows involving pre-processing, enhancement, and transcoding.

\subsection{UGC-VQA methods}

There are two main streams for user-generated content video quality assessment (UGC-VQA)~\cite{HVSvist,V-BLINDS,VQT,simpleVQA,Dover,pamifastvqa,tpqi,LSCT,robustVQA,contraiVideo,starVQA,Self_Supervised}. The first comprises traditional methods~\cite{V-BLINDS,TLVQM,VIIDEO,STEM}, which are constrained by the limitations of handcrafted features and lack of the adaptability to handle more complex UGC databases.
With the advancement of deep learning, the second stream learning-based methods often enable superior performance, which can be categorized into three main types: temporal fusion, multi-priors fusion, and fragment extraction. Temporal fusion-based methods~\cite{VSFA,discovqa,VQT,GSTVQA} aim to adaptively fuse quality features in the temporal domain. Multi-priors based methods~\cite{simpleVQA,BVQA,MDVQA,ADAQA} typically incorporate multi-priors into quality-aware features for final regression. Fragment-based methods~\cite{FastVQA, Dover} extract texture-level information and eliminate substantial spatio-temporal redundancies. 
However, above methods do not incorporate the ability of content-distortion understanding into the feature extraction process, which hinders their capability to address the two challenges in short-form video platforms.

%% file: sec/3_database.tex
\section{Our proposed KVQ Database}
\label{sec:kvq_dataset}
\tcr{To advance the progress of short-form video quality assessment, we built the first large-scale KVQ database, intending to assist the algorithm development. In contrast to traditional UGC VQA databases~\cite{CVD2014, KoNViD-1k,LSVQ,youtubeUGC}, our KVQ database exhibits the following distinctive features and advantages: (i) special but crucial application scenario, \ieno, short-form video platform, (ii) advancing content creation/generation modes and kaleidoscope contents, 
 (iii) practical and sophisticated processing workflows, (iv) unique scoring strategy, \ieno, the combination of absolute and ranking quality score. In the following sections, we will clarify the above features/advantages in detail.}
\subsection{Dataset Collection}
\label{sec Dataset Collection}
\tcr{Our dataset is composed of 4200 S-UGC videos, which is collected following two principles: (i) ensure the content diversity and distortion diversity as much as possible and (ii) satisfy the practical online statistics and application/requirements in the popular short-form video platforms. The pipeline of our dataset collection is shown in Fig.~\ref{fig:framework}. Notably, in practical application, the previous UGC-VQA methods usually perform poorly for content generated with advancing creation modes, such as special effects. Considering that, we collect the datasets from several typical creation modes, including three-stage, special effects, subtitled, live modes (Please see Fig.~\ref{fig:first}), and other traditional creation modes. The data are composed of nine primary content scenarios in the practical short-form video platform, including landscape, crowd, person, food, portrait, computer graphic (termed as CG), caption, and stage. In this way, these original user-uploaded data contents cover almost all existing creation modes and scenarios, and the ratio of each category of content satisfies the practical online statistics. To further align the video features in the practical platform, we make fine-grained video content adjustments based on typical 6 video features, \ieno, sharpness, complexity, blurriness, noise, blocky, and colorfulness. Based on the above collection strategies, we collect 600 original user-uploaded S-UGC videos for next-stage processing. 
}

\tcr{Most UGC databases, \egno, UGC-VIDEO~\cite{UGC-VIDEO}, simulate the video processing pipeline for UGC videos with single or simple processing tools, such as transcoding. However, in practical short-form video platforms, the video processing pipeline is sophisticated, including different pre-processing, transcoding, and enhancement tools, intending to enhance the subjective quality and reduce the coding bitrate. Moreover, the video processing pipeline is adaptive for each video based on its content and quality. Therefore, to build an applicable database, we exploit the representative video processing strategy in a practical short-form video platform for our KVQ database, which is shown in Fig.~\ref{fig:framework}, where enhancement $\phi_e(\cdot)$, pre-processing $\phi_p(\cdot)$, and transcoding $\phi_t(\cdot)$ work in a cascaded manner. Concretely, 50\% of high-quality videos are processed with six transcoding modes, since they do not need enhancement and pre-processing. Another 50\% of low-quality videos select one enhancement tool from tool pools of de-artifacts, denoise, and deblur. Then the pre-processing is made with a probability of 0.5 for enhanced low-quality data, followed by transcoding. In this way, 3600 processed S-UGC videos are obtained, which can be divided into three groups corresponding to three typical working flows, \ieno, $\phi_t(\cdot)$, $\phi_t(\phi_e(\cdot))$ and $\phi_t(\phi_p(\phi_e(\cdot)))$. Based on the above collection strategy, we collect 4200 S-UGC videos as our database. } 
\tcb{No questions on licenses existed in this work since the data collection is authorized by the short-form video platform and owners.}
\vspace{-1mm}
\subsection{Human Study}
\tcr{The human study is carried out with 15 professional researchers specializing in image processing in the standard environment for quality assessment. Despite the professional labeling, it is still hard to achieve fine-grained absolute scoring with single-stimulus (SS) methods~\cite{installations1999subjective}. To enable the fine-grained evaluation capability, we propose mixed scoring, where the absolute Mean Opinion Score (MOS) value is provided for each video with the range of \tcb{[1-5]} and the interval of 0.5, and the ranking score is provided for the indistinguishable S-UGC videos. For the absolute MOS value, we follow the standard subjective procedure in ITU-R BT 500.13~\cite{bt2002methodology}. Each participant is given the training with unified instruction. After scoring, the data cleaning process is performed for each video.
}

\tcr{We notice that there are two representative indistinguishable scenarios. The first scenario occurs for different video contents (\ieno,  non-homogeneous video pairs), where the difference of MOSs is less than 0.5. Another scenario is that the transcoding levels do not match their assessed quality order for the same content (\ieno, the homogeneous video pairs) since the adaptive enhancement and preprocessing. Therefore, to improve the fine-grained evaluation capability, we select 250 homogeneous video pairs and 250 non-homogeneous video pairs for ranking labeling.}

\begin{figure}
    \centering
    \includegraphics[width=0.45\textwidth]{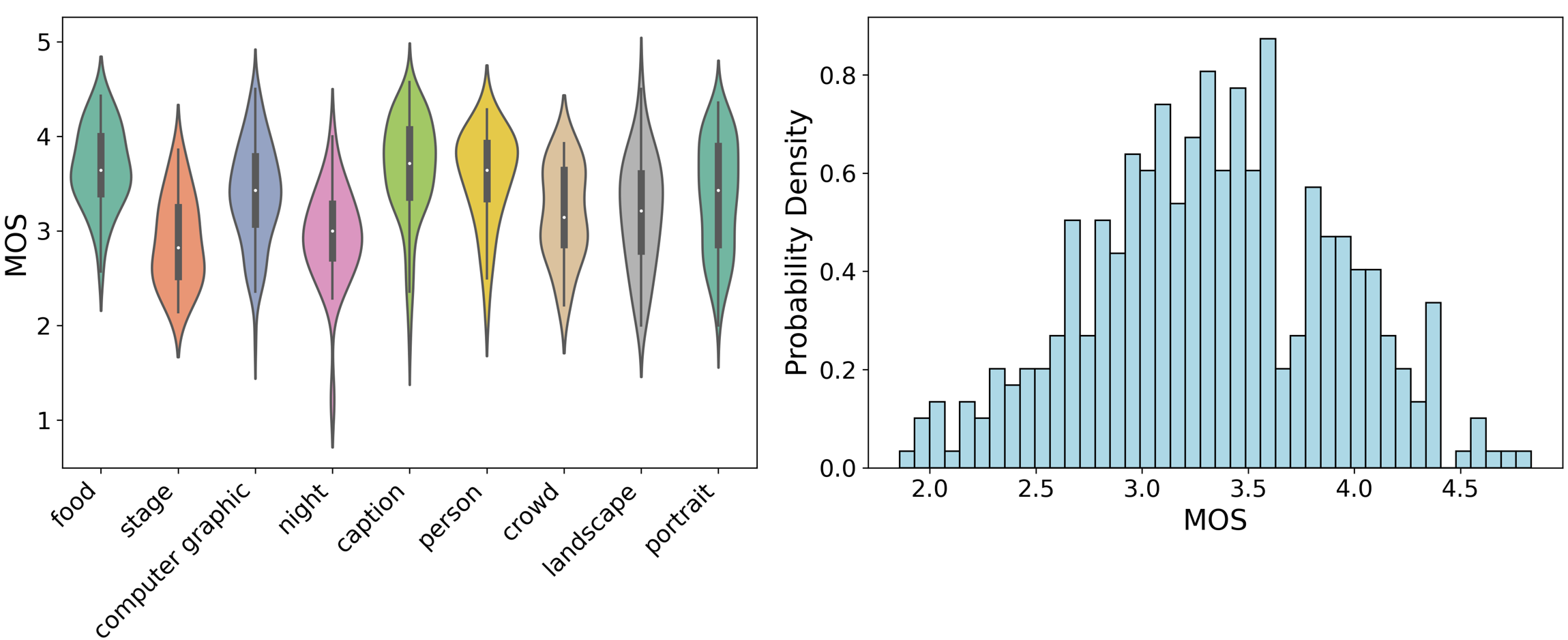}
    \caption{The MOS distribution of different semantic categories (a) and the histogram of the overall MOS distribution (b). }  
    \label{fig: class_distri}
\end{figure}

\begin{figure*}
    \centering
    \includegraphics[width=0.9\textwidth]{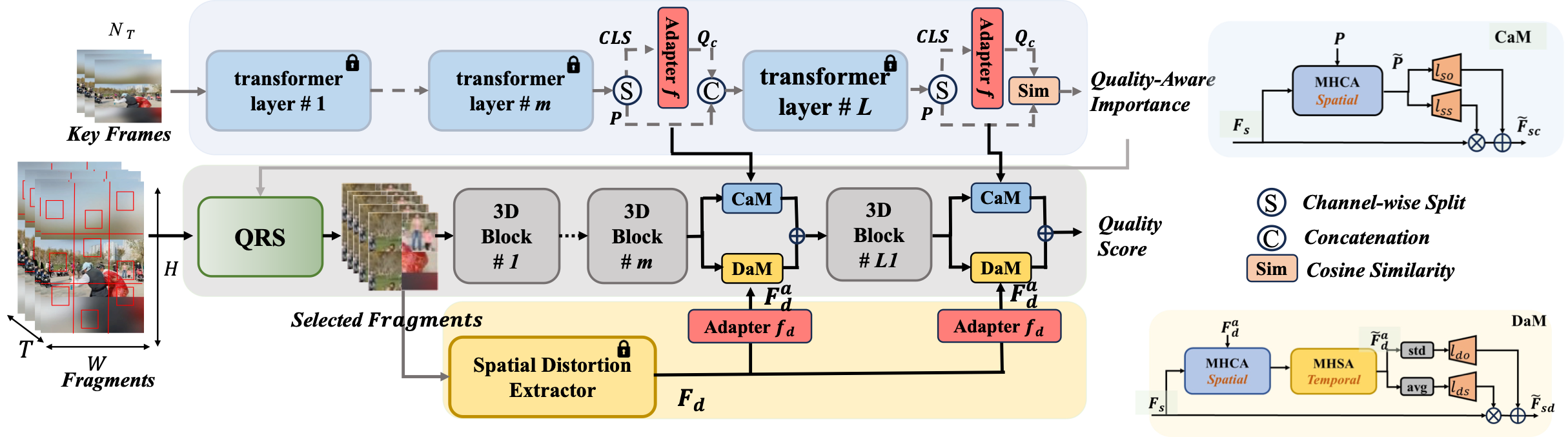}
    \caption{The overall framework of Kaleidoscope Short-form UGC Video Quality Evaluator (KSVQE). It contains  quality-aware region selection module (QRS) and content-adaptive modulation (CaM) to incorporating content understanding, and distortion-aware modulation (DaM) to enhance distortion understanding. } 
    \label{fig:ksvqe}
\end{figure*}
\subsection{Subjective Quality Analysis}
\tcr{In this subsection, we conduct a thorough analysis of the subjective quality score for our KVQ. Specifically, we visualize the MOS distribution for 9 content scenarios in Fig.~\ref{fig: class_distri}. We can observe that the MOS distributions of different contents are similar except for the night and stage scenarios,} \tcb{due to that the dark night scenario and complex stage motion are prone to cause a bad perception experience.}

\begin{figure}
    \centering
    \includegraphics[width=0.8\linewidth]{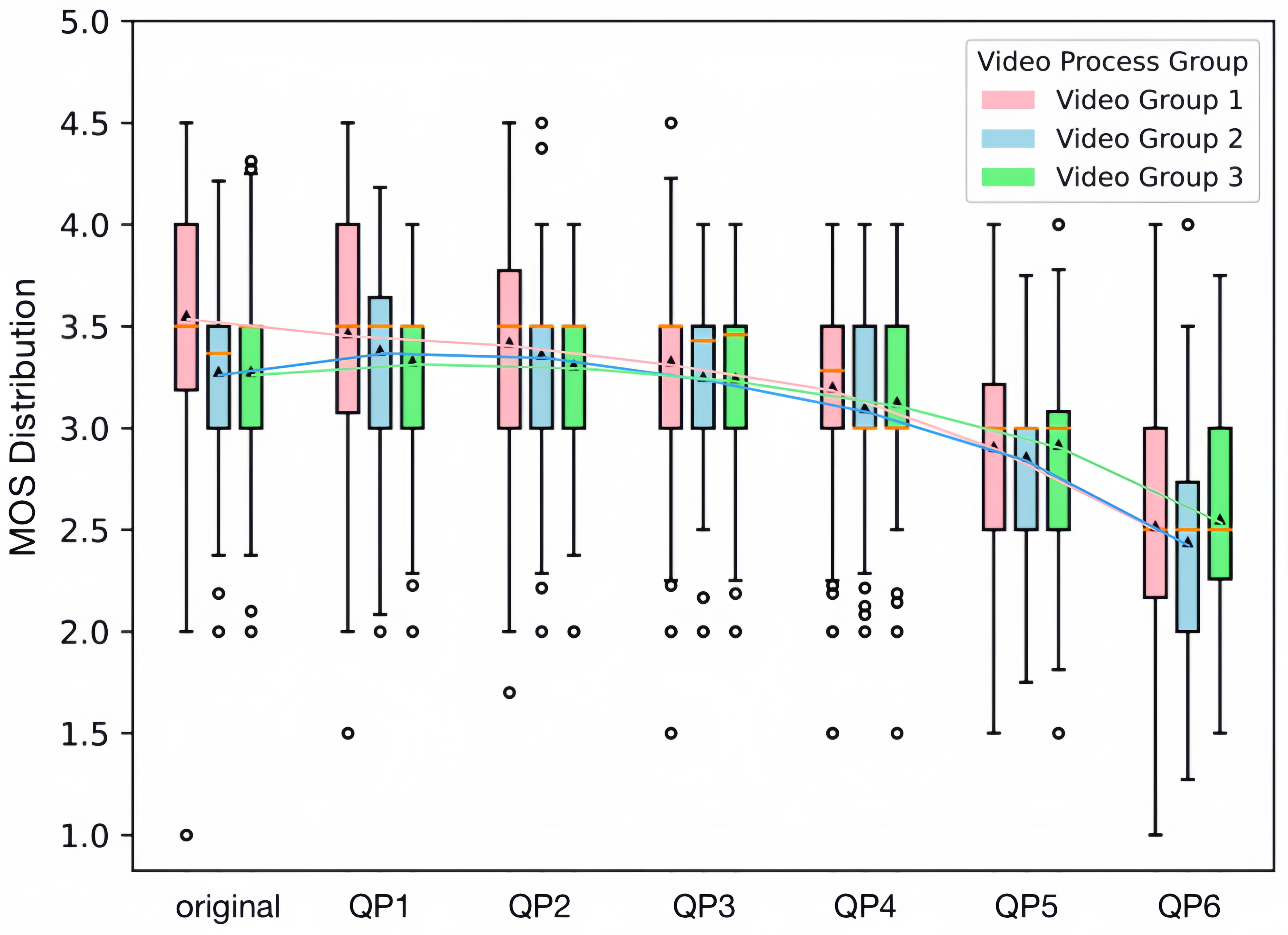}
    \caption{MOS distribution of videos of the three video groups corresponding to the three video processing workflows. }
    \label{fig:dist_score}
\end{figure}

\tcr{To investigate the impacts of different processing workflows on subjective quality, we visualize the MOS distribution of three video groups. As stated in section~\ref{sec Dataset Collection}, based on the distortions in 600 original S-UGC videos, we can divide it into three groups, where the high-quality video group 1 is only processed with different transcoding modes. From Fig.~\ref{fig:dist_score}, we can observe that the subjective quality will decrease with the QP increases since the compression artifacts increase. By comparing the subjective quality of original videos and the processed ones in the first and second QP intervals in Video Group 2 (\ieno, processed with enhancement and transcoding), we can find that the enhancement tools can improve the subjective quality effectively despite the compression occurring. Since the pre-processing is achieved with a probability of 0.5, the comparison between Video Group 2 and 3 has demonstrated that the pre-processing can eliminate the decrease of subjective quality, especially in the low-bitrate range, such as QP interval six.} 

\tcr{The above subjective quality analysis are consistent well with the functions of different video processing tools, which proves the reliability of our human study in some content. }
\tcb{More details will be provide in the \textbf{Appendix}. }

%% file: sec/4_method.tex
\section{Our Proposed Method}
\label{sec:proposed method}
\tcr{To solve two primary challenges in the S-UGC quality assessment: (i) the presence of advanced creation modes and kaleidoscope contents prevent the UGC VQA model from identifying the quality-determined regions, (ii) the sophisticated processing flows increase the difficulties for distortion distinguishment, we propose the first short-form UGC video quality evaluator, \ieno, KSVQE. The purpose is to enhance the content and distortion understanding capability for the VQA model under special S-UGC scenarios and eliminate the intervention from quality-unrelated regions. The whole framework of KSVQE is shown in Fig.~\ref{fig:ksvqe}. We adopt the powerful 3D-Swin Transformer as our backbone for quality regression, and its effectiveness has been validated in a series of works~\cite{FastVQA,NTIRE1,discovqa,zoomVQA}. To improve the training efficiency while keeping the diversity of contents, we exploit the fragment strategy~\cite{FastVQA, Dover} for the $T$ frames of S-UGC video $X$, which divides the original video into $N$ patches and randomly samples a fragment with the size $h\times w$ from each patch. Then the composite image with $N$ fragments is utilized as the input $\Tilde{X} \in \mathbb{R}^{T\times Nh \times Nw}$ of our KSVQE.}

\subsection{Content Understanding}
\label{sec:Content Understanding}
\tcr{It is noteworthy that not all patches in one S-UGC video are quality-related and there are some patches in one image that even intervene in the evaluation of quality since the new creation modes, such as special effects in Fig.~\ref{fig:first}. To solve this, we propose to increase the content understanding capability of our KSVQE with the help of a pre-trained large foundation model CLIP~\cite{CLIPsemantic,CLIPsemantic2,CLIPsemantic1,CLIPsemantic3,CLIPsemantic4,li2023graphadapter}.} \tcb{In order to mine the global semantic, We feed $N_t$ key frames sampled from the resized video into CLIP visual encoder.}

\tcr{CLIP has revealed its powerful fine-grained semantic perception capability~\cite{CLIPsemantic5,CLIPsemantic6,CLIPsemantic7}, attributed to millions of training text-image pairs from the web. Despite that, directly applying it for content understanding still does not meet the requirements in video quality assessment since it is required to be quality-aware. Therefore, we propose the quality adapter, which is incorporated into the class token (\ieno, semantics) of the last two layers of CLIP to achieve patch-wise quality-aware content understanding with the constraint of quality assessment. Concretely, given the output features $[CLS, P]$ of CLIP for key frame in S-UGC video, where $CLS$ is the class token and $P\in \mathbb{R}^{N\times C_c}$ is the features corresponding each patch. With the quality adapter~\cite{muNet,Convadapter} $f(\cdot)$, the semantic class token can be adapted into quality-aware space as $Q_c = f(CLS)$. And the patch-wise quality-aware semantic importance $I$ can be computed with
\begin{equation}
   I =Q_{c} P^{T}/\|Q_{c}\| \|P^{T}\|
\end{equation}} 
\tcr{Based on this, we propose two innovative modules to eliminate the impacts of advancing creation modes and improve the adaptation capability for the quality assessment of different kaleidoscope contents. }

\vspace{-2mm}
\paragraph{Quality-aware Region Selection (QRS).}
\tcr{It is noteworthy that the advancing creation models usually introduce amounts of quality-unrelated content that is ambiguous for VQA. For instance, in the three-stage creation, large-area backgrounds with special effects are not focused by humans. To mitigate the side impacts of quality-unrelated regions, we propose to achieve the quality-aware region selection. Specifically, based on quality-aware semantic importance score $I\in \mathbb{R}^{N}$ obtained from adapted CLIP, we select the most quality-aware $K$ patches by their indexes as $inds = \mathrm{TopK}(I)$. Here the Differentiable TopK Selection~\cite{cordonnier2021differentiable,stts} is exploited to enable the training of QRS.}

\vspace{-2mm}
\paragraph{Content-adaptive Modulation (CaM).}
\label{sec: Semantic Modulation}
\tcr{Notably, the perception of the image is closely associated with the content semantics~\cite{QPT,REIQA,BlueSky,zhou2023blind,TTLIQA,SemanticIQA} (\ieno, semantic-adaptive). To boost the semantic adaptability of our KSVQE, we propose content-adaptive modulation (CaM) in Fig.~\ref{fig:ksvqe}, where the fine-grained semantics of each patch $P$ from CLIP from the last two layers are inserted into the same position of 3D swin transformer. 
Concretely, we utilize the features $F_s$, including selected $K$ patches, from 3D swin transformer as the query, to warp the related semantics  $\Tilde{P}$  from corresponding CLIP features $P$ with multi-head cross attention (MHCA).
Then the spatial-wise scale and offset modulation coefficients $\gamma_s$ and $\beta_s$ are generated with convolution layers $l_{ss}$ and $l_{so}$ to fuse the fine-grained semantic information of each patch. 
\begin{equation}
    \Tilde{F}_{sc} = l_{ss}(\Tilde{P})F_s+l_{so}(\Tilde{P}).
\end{equation}
}
\vspace{-4mm}
\subsection{Distortion Understanding and Modulation}
\label{sec:Distortion Understanding}
\tcr{To improve the quality assessment for S-UGC videos, it is crucial to enhance the distortion understanding capability for the quality evaluator \tcb{, which can handle the challenge of indistinguishability of distortions arising from the sophisticated process workflows in Sec.\ref{sec Dataset Collection}}. }

\tcr{To achieve this, we need to extract the distortion priors from selected S-UGC fragments and fuse them into the quality evaluator. In this work, we exploit the popular pre-trained CONTRIQUE~\cite{CONTRIQUE} as a spatial distortion extractor to extract fragment-wise features as $F_d = \mathcal{D}(\Tilde{X})$. However, it inevitably suffers from the distribution shifts since the distortions in our KVQ database are greatly different from existing databases. Considering that, we propose the distortion adapter $f_d$ to adapt the pre-trained CONTRIQUE to the target distortion distribution in our KVQ database with the distortion contrastive loss, where distortions in the same processing pattern are regarded as positive pairs and others as negative pairs. In this way, we can obtain a good spatial-wise distortion prior for S-UGC videos as $F_d^a=f_d(F_d)$. }

\paragraph{Distortion-aware modulation (DaM)}. To incorporate the distortion prior into our KSVQE, we propose the distortion-aware modulation by exploiting the multi-head cross attention (MHCA) to warp the captured spatial distortion features with the query of the quality feature $F_s$ from 3D swin transformer. To let the evaluator perceive the temporal distortion in S-UGC videos, we also exploit the multi-head self-attention (MHSA) to interact with the warped spatial distortion features as $\Tilde{F}_d^a=\mathrm{MHSA}(\mathrm{MHCA}(F_s, F_d^a)$. As shown in Fig.~\ref{fig:ksvqe}(c), the distortion modulation is achieved with the channel-wise feature style modulation~\cite{liu2021discovering}, and the channel-wise scale and offset are obtained \tcb{ through applying two linear layers (\ieno, $l_{ds}$ and $l_{do}$) to the mean and standard deviation of feature $\Tilde{F}_d^a $}:
\tcb{\begin{equation}
    \Tilde{F}_{sd}  = l_{ds}(\mathrm{std}(\Tilde{F}_d^a))F_s + l_{do}(\mathrm{avg}(\Tilde{F}_d^a)).
\end{equation}}

With the cooperation of QRS, CaM, and DaM, our KSVQE performs greatly on the quality assessment of S-UGC videos, which bridges the void in the short-form video quality assessment.

%% file: sec/5_experiment.tex
\section{Experiment}

\begin{table*}[t]

  \centering
  \footnotesize
  \caption{Performance of existing SOTA methods and the proposed KSVQE on our built KVQ and four in-the-wild VQA datasets. The ``N/A" means missing corresponding results in the original paper. The best and second-best results are \textbf{bolded} and \underline{underlined}.}
 \label{tab:compare}
  \begin{tabular}{c|cc|cc|cc|cc|cc|cc}
    \toprule
    \multirow{2}{*}{Method} & \multicolumn{2}{c|}{KVQ} & \multicolumn{2}{c|}{KoNViD-1k} & \multicolumn{2}{c|}{YouTube-UGC} & \multicolumn{2}{c|}{LIVE-VQC} & \multicolumn{2}{c|}{LSVQ\_test} & \multicolumn{2}{c}{LSVQ\_1080p} \\
    & SROCC & PLCC & SROCC & PLCC & SROCC& PLCC & SROCC & PLCC & SROCC & PLCC & SROCC & PLCC \\ 
    \midrule
    VIQE~\cite{VIQE} &0.221&0.397 &0.628&0.638
     &0.513 &0.476  &0.659 &0.694 &N/A &N/A &N/A &N/A \\
    TLVQM~\cite{TLVQM}  &  0.490& 0.509  & 0.773  & 0.768    & 0.669  & 0.659 &0.798 & 0.802& 0.772  & 0.774   & 0.589  & 0.616 \\
    RAPIQUE~\cite{RAPIQUE} &0.740&0.717&0.803&0.817 & 0.759&0.768 &0.754&0.786 &N/A&N/A &N/A &N/A
    \\
    VIDEVAL~\cite{UGC-VQA}   & 0.369   & 0.639 & 0.773& 0.768& 0.669& 0.659& 0.752&0.751 & 0.795& 0.783& 0.545& 0.554          \\ \midrule
    VSFA~\cite{VSFA}  &  0.762  &  0.765 & 0.773& 0.775& 0.724& 0.743&0.773&0.795 & 0.801& 0.796& 0.675& 0.704          \\
    GSTVQA~\cite{GSTVQA}  & 0.786 & 0.781 & 0.814& 0.825 & N/A  & N/A &0.788&0.796 & N/A         & N/A &N/A &   N/A       \\
    PVQ~\cite{PVQ} &0.794&0.801 & 0.791 & 0.786 &N/A &N/A &  0.827&0.837  & 0.827&0.828 &0.711&0.739 \\ 
    SimpleVQA~\cite{simpleVQA} & \underline{0.840}  & \underline{0.847} & 0.856& 0.860  & 0.847          & 0.856 & N/A&N/A & 0.867 & 0.861 & 0.764 & 0.803          \\ 
    FastVQA~\cite{FastVQA}   &0.832  &0.834   & 0.891 & 0.892 & \underline{0.855}          & \underline{0.852}&\underline{0.849} &0.862  & 0.876  & 0.877 & \underline{0.779}  & \underline{0.814}          \\ 
    Dover*~\cite{Dover}\tablefootnote{Exclude the aesthetic branch for a fair comparison~\cite{Dover}.}   &0.833  &0.837   &\underline{ 0.908} & \underline{0.910}  & 0.841& 0.851&0.844& \underline{0.875} &\underline{0.877}  &\underline{0.878} &0.778  &0.812         \\ \midrule
   
KSVQE & \textbf{0.867} & \textbf{0.869} & \textbf{0.922} & \textbf{0.921} & \textbf{0.900} & \textbf{0.912} &\textbf{0.861}&\textbf{0.883} & \textbf{0.886} & \textbf{0.888} & \textbf{0.790} & \textbf{0.823}  \\
    \bottomrule
  \end{tabular}
  \label{tab:sota}
\end{table*}

\vspace{-1mm}
\begin{table*}[]
\caption{Ablation study for the three proposed components (\ie, QRS, CaM and DaM).}
\label{tab:main_abl}

 \centering
  \footnotesize
\begin{tabular}{cc|c|cc|cc|cc}

\toprule

\multicolumn{2}{c|}{Content Understanding}                   & \multicolumn{1}{c|}{Distortion Understanding}  & \multicolumn{2}{c|}{KVQ}         & \multicolumn{2}{c|}{KoNViD-1k}       & \multicolumn{2}{c}{YouTube-UGC}    \\ 
\multicolumn{1}{c}{QRS} &CaM & \multicolumn{1}{c|}{DaM}                             & \multicolumn{1}{c}{SROCC} & PLCC & \multicolumn{1}{c}{SROCC} & PLCC & \multicolumn{1}{c}{SROCC} & PLCC \\ 
\midrule
\multicolumn{1}{c}{\CheckmarkBold}            &{\CheckmarkBold}              & {\CheckmarkBold}                                                & \multicolumn{1}{c}{\textbf{0.867}}      & \textbf{0.869}     & \multicolumn{1}{c}{\textbf{0.922} }      & \textbf{0.921}     & \multicolumn{1}{c}{\textbf{0.900}}      &\textbf{0.912}      \\ 
\midrule
\multicolumn{1}{c}{\XSolidBrush}             & {\XSolidBrush}               & {\XSolidBrush}                                                 & \multicolumn{1}{c}{0.832}      &0.834      & \multicolumn{1}{c}{0.891}      &0.892      & \multicolumn{1}{c}{0.855 }      & 0.852     \\ 
\multicolumn{1}{c}{\CheckmarkBold}             & {\XSolidBrush}               & {\XSolidBrush}                                                 & \multicolumn{1}{c}{0.847}      &0.853     & \multicolumn{1}{c}{0.917}      & 0.920     & \multicolumn{1}{c}{0.887}      &0.903      \\ 

\multicolumn{1}{c}{\XSolidBrush}            & {\CheckmarkBold}               & {\XSolidBrush}                                                  & \multicolumn{1}{c}{0.842}      &0.847      & \multicolumn{1}{c}{0.918}      &0.914      & \multicolumn{1}{c}{0.895}      &0.900      \\  

\multicolumn{1}{c}{\XSolidBrush}             & {\XSolidBrush}             & {\CheckmarkBold}                                                  & \multicolumn{1}{c}{0.839}      & 0.843     & \multicolumn{1}{c}{0.915}      & 0.914     & \multicolumn{1}{c}{0.893}      & 0.910     \\ 
\multicolumn{1}{c}{\XSolidBrush}            &{\CheckmarkBold}              & {\CheckmarkBold}                                                  & \multicolumn{1}{c}{0.850}      &0.852     & \multicolumn{1}{c}{0.921}      &0.912      & \multicolumn{1}{c}{0.896}      &0.904      \\ 

\bottomrule
\end{tabular}
\end{table*}

\begin{table}[]
\setlength{\tabcolsep}{0.4mm}{

\caption{Performance on the ranking of pairs in the KVQ dataset. There are a total of 100 pairs, comprising 50 non-homogeneous pairs and 50 homogeneous pairs.}
\label{tab:rank}

\footnotesize
 \begin{center}
 \vspace{-3mm}
\begin{tabular}{c|c|c|c}
\toprule
Rank      & non-homogeneous & homogeneous & all pairs\\ \midrule

TLVQM~\cite{TLVQM}     & 0.56         &0.64       & 0.6    \\ 
VIDEVAL~\cite{UGC-VQA}   &  0.36          & 0.60       & 0.48    \\ \midrule
VSFA ~\cite{VSFA}     & 0.54           & 0.92       &  0.73   \\ 
GSTVQA ~\cite{GSTVQA}      & 0.58           &\textbf{0.98}        &  0.78   \\ 
SimpleVQA~\cite{simpleVQA}   & 0.58           & 0.96       & 0.77    \\ \
FastVQA~\cite{FastVQA}   &0.66            & 0.86      & 0.76    \\ 

Dover* ~\cite{Dover}  &  0.70        & 0.88       & 0.79    \\ \midrule
KSVQE     &\textbf{0.76}            & 0.86      &  \textbf{0.81}   \\  \bottomrule
\end{tabular} \end{center}}
\vspace{-3mm}
\end{table}

\subsection{UGC-VQA Databases}
We verify our framework on four datasets: our proposed KVQ dataset, KoNViD-1k~\cite{KoNViD-1k}, Youtube-UGC~\cite{youtubeUGC}, \tct{LIVE-VQC~\cite{LIVE-VQC}} and LSVQ~\cite{LSVQ}. For the KVQ dataset, we randomly split it into an 80\% training set and a 20\% test set according to the reference content. For LSVQ, we follow the public split version ~\cite{LSVQ} to validate our method. For the rest of the databases, we follow the previous standard method~\cite{FastVQA,unfiedVQA} and split the databases with an 80\%-20\% train-test ratio. And the performance reported depends on the checkpoint of the last iteration in training.

\subsection{Implementation Details}

For the details about KSVQE, we utilize the CLIP visual encoder from ViT-B~\cite{CLIP} to extract semantic priors.
For KSVQE, the input fragments are of size $32 \times 288 \times 288$ with a 2-frame interval, consisting of $(9\times9)$ fragments, each of size 32. After region selection of QRS, the input for 3D Swin Transformer is realistically reshaped to $32 \times 224 \times 224$ with $(7\times7)$ fragments. For visual ViT-B~\cite{CLIP} of CLIP, we resize the original video in the spatial dimension to be $32 \times 224 \times 224$. Regarding CONTRIQUE, we feed each fragment with a size of $32 \times 32$. \tct{And the number of CLIP layers used for modulation is set as 2 through our optimal experiments results.}
We adopt two widely used criteria for performance evaluation: Pearson linear correlation coefficient (PLCC) and Spearman rank order correlation coefficient (SROCC). A higher value for these coefficients indicates a stronger correlation with quality annotations. 
Following~\cite{FastVQA,compressedVQA,li2023freqalign,liu2022sourceBIQA}, we apply PLCC loss for gradient descent to optimize KSVQE. More training details can be found in the \textbf{Appendix}.
\vspace{-1mm}
\subsection{Experiment Results}

To verify the effectiveness of our proposed \ours, We select seven UGC-VQA methods for comparison: traditional-based methods (VIQE~\cite{VIQE}, TLVQM~\cite{TLVQM}, RAPIQUE~\cite{RAPIQUE} and VIDEVAL~\cite{UGC-VQA}), deep learning-based methods (VSFA~\cite{VSFA}, GSTVQA~\cite{GSTVQA},PVQ~\cite{PVQ},  SimpleVQA~\cite{simpleVQA}, FastVQA~\cite{FastVQA} and Dover~\cite{Dover}). \tct{For fair comparison without pretrained weight on KVQ, we remove the aesthetic
branch for Dover as Dover* due to the lack of aesthetic scores.} As shown in Table~\ref{tab:compare}, the traditional TLVQM and VIDEVAL that rely on manual feature extraction face challenges in addressing complex UGC-VQA scenarios.
Specifically, our proposed KSVQE demonstrates superior performance across the KVQ, KoNViD-1k, and Youtube-UGC datasets. Notably, KSVQE outperforms the second-best method Dover* (w.o. aesthetic branch) by a substantial margin of 0.034/0.032 in terms of SROCC and PLCC on KVQ, 0.014/0.011 on KoNViD-1k, and 0.059/0.061 on Youtube-UGC, 0.009/0.010 on LSVQ.
It illustrates that with the help of content and distortion understanding, KSVQE can achieve accurate quality perception.

We also test KSVQE and multiple SOTA VQA methods on the ranked pair in Table~\ref{tab:rank}. And the accuracy is used to evaluate the performance of rank-pair prediction. From the results, it can be seen that our KSVQE can exceed the second-best methods Dover* with 0.06 of accuracy in non-homogeneous video pairs and 0.02 in all video pairs. It is evident that distinguishing quality in non-homogeneous video pairs is more challenging compared to homologous video pairs. This aligns with the difficulty arising from the presence of various creative modes, kaleidoscopic content scenarios, and indistinguishable distortions of sophisticated workflows. The accurate identification of quality becomes inherently difficult in such scenarios.

\tct{As for cross-dataset evaluation, we conduct two cross-dataset evaluations: ``KVQ$\rightarrow$other datasets" and ``other datasets$\rightarrow$KVQ" in Table~\ref{tab:cross2} and Table~\ref{tab:cross1}. We can find that: i) In above two settings, KSVQE can obtain the optimal performance, which shows that KSVQE outperforms other methods in generalization. ii) Through comparing the generalization performances of ``KVQ$\rightarrow$other datasets" and ``other datasets$\rightarrow$KVQ", our KVQ is more challenging than others, since training on KVQ yields good results on other datasets, while the reverse is worse. }

\begin{table}

\centering

\caption{The cross-dataset evaluations of ``other datasets $\rightarrow$ KVQ".}
\vspace{-2mm} 
\centering

\resizebox{\linewidth}{!}{\begin{tabular}{c|c|c|c}
   \toprule
Test:KVQ & {KoNViD-1k} & {YouTtbe-UGC}  & {LIVE-VQC}\\   
     \midrule
   SimpleVQA  &0.459/0.394&0.396/0.401& 0.345/ 0.392 \\
   FastVQA   &0.506/0.480  &0.450/0.409 &0.505/0.496    \\
   KSVQE  & \textbf{0.528/0.504}& \textbf{0.499/0.412 }&\textbf{0.539/0.533}  \\
    \bottomrule
  \end{tabular}
  } 
  \label{tab:cross1}
\end{table}

\begin{table}
\vspace{-2mm}

\caption{The cross-dataset evaluations of ``KVQ $\rightarrow$ other datasets".}
 \vspace{-2mm}
\resizebox{\linewidth}{!}{\begin{tabular}{c|c|c|c}
   \toprule
 { Train:KVQ}  & {KoNViD-1k} & {YouTtbe-UGC}  & {LIVE-VQC} \\   
     \midrule
   SimpleVQA &0.475/0.481   & 0.675/0.674  & 0.528/0.521\\
   FastVQA    &0.641/0.654 & 0.645/0.676  & 0.614/0.675     \\ 
  
KSVQE  &\textbf{0.650/0.661} & \textbf{0.742/0.764}&\textbf{0.720/0.768}\\ \bottomrule
  \end{tabular}}
  \label{tab:cross2}
  \end{table}
\subsection{Ablation Study}

To validate the effectiveness of the three core components: Quality-aware Region Selection module (QRS), Content-adaptive Modulation (CaM), and Distortion-aware Modulation (DaM) in our KSVQE, a comprehensive ablation study is conducted and the results are presented in Table~\ref{tab:main_abl}. In Table~\ref{tab:main_abl}, the $3^{th}$ row, which does not include the three components, serves as our baseline: a fully-trained 3D Swin Transformer with fragment input of dimensions $32\times224\times224$. Meanwhile, the $2^{th}$ row, which incorporates all three components, represents our proposed method, KSVQE.
\vspace{-2mm}
 \paragraph{The effectiveness of QRS.} By comparing the $3^{th}$ and $4^{th}$ rows, the inclusion of QRS yields a significant improvement of 0.015/0.019 over the baseline on the KVQ database. 
This observation highlights the advantage of identifying
the quality-determined region and dropping the quality-unrelated contents, such as large areas of solid color background in three-stage short videos, in the assessment of short video quality. The visualization result of selected fragments can be seen in the \textbf{Appendix}.
In order to prove the fine-grained semantic understanding of CLIP can mine the quality-determined region, we replace the learnable selection with some variants, and the experiment results can be seen in Table~\ref{tab:QRS}. The ``baseline" denotes the original fragment sampling and the ``RS" represents the random selection of a region from the region candidate of interest. We can see that our QRS achieves the best performance compared with these two variants, which shows the effectiveness of mining the quality-determined region.
\vspace{-3mm}
 \paragraph{The effectiveness of CaM.} From the results of $3^{th}$ row and $5^{th}$ row in Table~\ref{tab:main_abl}, we can observe that the CaM can bring the performance gain on SROCC/PLCC compared with the baseline in all databases. Especially for KVQ, the CaM exhibits a performance improvement of 0.010/0.013 on SROCC and PLCC, which illustrates the necessity of quality perception associated with different semantics for short-form videos with various generation modes.
In Table~\ref{tab:CaMDaM}, we present various modulation variants. The first variant, ``CA", involves only multi-head cross attention. The second variant, ``SM" retains only spatial-wise modulation. The Third variant, ``CA+CM" combines multi-head cross attention and channel-wise modulation.
Through the comparison of these variants of CaM in Table~\ref{tab:CaMDaM}, it is demonstrated that the ``CA+SM" (\ieno, our CaM) provide the best performance, allowing for a richer quality-aware semantic instruction in KSVQE on the spatial dimension.

\begin{table}[]
\setlength{\tabcolsep}{0.3mm}{
\caption{Ablation study for multiple variants of selection in QRS.}
\vspace{-0.3cm}
\label{tab:QRS}
\footnotesize
 \begin{center}
\begin{tabular}{c|cc|cc|cc}
\toprule
\multicolumn{1}{c|}{\multirow{2}{*}{Region Selection}} & \multicolumn{2}{c|}{KVQ}         & \multicolumn{2}{c|}{KoNViD-1k}       & \multicolumn{2}{c}{YouTube-UGC}    \\ 
\multicolumn{1}{c|}{}                                 & \multicolumn{1}{c}{SROCC} & PLCC & \multicolumn{1}{c}{SROCC} & PLCC & \multicolumn{1}{c}{SROCC} & PLCC \\ \midrule
baseline                                                  & \multicolumn{1}{c}{0.832}      &  0.834    & \multicolumn{1}{c}{0.907 }      & 0.909     & \multicolumn{1}{c}{0.884 }     &   0.900   \\
RS                                   & \multicolumn{1}{c}{0.843}      &0.845      & \multicolumn{1}{c}{0.916}      &0.915      & \multicolumn{1}{c}{\textbf{0.896}  }      &  0.905    \\ \midrule
QRS                                           & \multicolumn{1}{c}{\textbf{0.847}}      &\textbf{0.853}      & \multicolumn{1}{c}{\textbf{0.917}  }      & \textbf{0.920}     & \multicolumn{1}{c}{0.894}      & \textbf{0.906}    \\ \bottomrule
\end{tabular} \end{center}}
\vspace{-3mm}
\end{table}

\vspace{-4mm}
\begin{table}[]
\footnotesize
\setlength{\tabcolsep}{0.08mm}{
\caption{Ablation study for multiple variants for CaM and DaM.}
\vspace{-0.3cm}
\label{tab:CaMDaM}
\vspace{-2mm}
 \begin{center}
\begin{tabular}{c|cc|cc|cc}
\toprule
\multicolumn{1}{c|}{\multirow{2}{*}{CaM}} & \multicolumn{2}{c|}{KVQ}         & \multicolumn{2}{c|}{KoNViD-1k}       & \multicolumn{2}{c}{YouTube-UGC}    \\ 
\multicolumn{1}{c|}{}                              & \multicolumn{1}{c}{SROCC} & PLCC & \multicolumn{1}{c}{SROCC} & PLCC & \multicolumn{1}{c}{SROCC} & PLCC \\ \midrule

CA                                         & \multicolumn{1}{c}{0.835}      & 0.831     & \multicolumn{1}{c}{0.912}      & 0.910     & \multicolumn{1}{c}{0.883 }      & 0.899     \\ 
SM                                        & \multicolumn{1}{c}{0.833}      & 0.840     & \multicolumn{1}{c}{0.915}      & 0.912     & \multicolumn{1}{c}{0.887 }      & 0.900     \\ 
CA+SM (CaM)                           & \multicolumn{1}{c}{\textbf{0.842}}      & \textbf{0.847}     & \multicolumn{1}{c}{\textbf{0.918}}       & \textbf{0.914}     & \multicolumn{1}{c}{\textbf{0.895}}      &\textbf{0.900}      \\ 
CA+CM                          & \multicolumn{1}{c}{0.838}      & 0.840     & \multicolumn{1}{c}{0.913}      & 0.916     & \multicolumn{1}{c}{0.879 }      & 0.893     \\ \midrule \midrule
\multicolumn{1}{c|}{\multirow{2}{*}{DaM}} & \multicolumn{2}{c|}{KVQ}         & \multicolumn{2}{c|}{KoNViD-1k}       & \multicolumn{2}{c}{YouTube-UGC}    \\ 
\multicolumn{1}{c|}{}                                & \multicolumn{1}{c}{SROCC} & PLCC & \multicolumn{1}{c}{SROCC} & PLCC & \multicolumn{1}{c}{SROCC} & PLCC \\ \midrule

CASA                                          & \multicolumn{1}{c}{0.830}      &0.837      & \multicolumn{1}{c}{0.911}      &0.909      & \multicolumn{1}{c}{0.887 }      & 0.898     \\ 
CM                                          & \multicolumn{1}{c}{0.831}      &0.836      & \multicolumn{1}{c}{0.909}      & 0.913     & \multicolumn{1}{c}{0.895}      &  0.904     \\ 
CASA+CM (DaM)                           & \multicolumn{1}{c}{\textbf{0.839}}      & \textbf{0.843}      & \multicolumn{1}{c}{\textbf{0.915}}      &\textbf{0.914}      & \multicolumn{1}{c}{\textbf{0.893}}      &  \textbf{0.910}    \\ 

CASA+SM                        & \multicolumn{1}{c}{0.830}      &0.833      & \multicolumn{1}{c}{0.913}      &0.914      & \multicolumn{1}{c}{0.894 }      & 0.903     \\ \bottomrule
\end{tabular}\end{center}}
\vspace{-0.3cm}
\end{table}

 \paragraph{The effectiveness of DaM.}
As depicted in Table~\ref{tab:main_abl}, we conducted a performance comparison between the results in the $6^{th}$ row (\ieno, ``baseline+DaM") and the baseline. Our findings revealed that DaM led to an improvement of 0.009 of PLCC on the KVQ database. This result underscores the significance of distortion guidance, particularly in distinguishing more complex distortions in sophisticated processing workflows. 
Similarly, we also evaluate the other variants for distortion modulation: only cross attention and self-attention (\ieno, ``CASA"), only channel-wise modulation (\ieno, ``CM") and a combination of attention modules and spatial-wise modulation(\ieno, ``CASA+SM"). We can find that ``CASA+CM" (\ieno, our DaM) achieves the best performance compared with these variants, which shows the necessity of the local distortion aggregation and channel-wise style injection of distortion prior.
\vspace{-3mm}
 \paragraph{The effectiveness of CaM and DaM.}
To analyze the effectiveness of the combination of CaM and DaM, we compare the $7^{th}$ row (\ieno, CaM+DaM) with the $5^{th}$ row (\ieno, CaM) and the $6^{th}$ row (\ieno, DaM) in Table~\ref{tab:main_abl}. The combination has a 0.008/0.005 increase in the terms of SROCC and PLCC compared with CaM and exhibits a performance improvement of 0.011/0.009 than DaM on the KVQ database. It reveals that the proposed modules can effectively ease the two primary challenges of kaleidoscope content and indistinguishable distortion in the KVQ database.
\vspace{-3mm}
\tct{\paragraph{The selection of content extractor and distortion extractor.}
To verify the effectiveness of enhanced CLIP (with adapter-style training) for quality-ware content mining, we replace CLIP with CLIPIQA+~\cite{CLIPIQA} and LIQE~\cite{LIQE} in Table~\ref{tab:CONA}. From the results, our KSVQE with enhanced CLIP can obtain the optimal correlation performance on KVQ, which shows the ability of the enhanced CLIP to capture quality-aware content. The visualization results can be seen in Appendix. As for the effectiveness of CONTRIQUE for distortion identification in KVQ, we choose GraphIQA~\cite{GraphIQA} or ReIQA~\cite{REIQA} to substitute CONTRIQUE in KSVQE, which is shown in Table~\ref{tab:DISA}. From these results, we can see that CONTRIQUE with distortion-aware contrastive learning can be adapted well to distortion space in KVQ.}

\begin{table}[]

\caption{Different selection for content extractor in KSVQE, in which ``XXX/XXX" represent ``SROCC/PLCC".}
\label{tab:CONA}

\footnotesize
\centering
\resizebox{\linewidth}{!}{\begin{tabular}{c|c|c|c}
   \toprule
{Model}  & {KVQ}  & {KoNViD} & {YT-UGC}   \\
     \midrule
   CLIPIQA+ &0.862/0.856   & 0.916/0.915 & 0.888/0.899
    \\ 
  LIQE   &0.848/0.854  &0.920/0.917 &0.892/0.896 
  \\  \midrule
  KSVQE &\textbf{0.867/0.869} &\textbf{0.922/0.921} & \textbf{0.900/0.912} \\
    
    \bottomrule
  \end{tabular}}

\end{table}

\vspace{-3mm}
\begin{table}[]

\caption{Different selection for distortion extractor in KSVQE, in which ``XXX/XXX" represent ``SROCC/PLCC".}
\label{tab:DISA}
\footnotesize
\centering
\resizebox{\linewidth}{!}{\begin{tabular}{c|c|c|c}
   \toprule
{Model}  & {KVQ}  & {KoNViD} & {YT-UGC}  \\
     \midrule
   ReIQA   &0.858/0.851  &0.921/0.921 & 0.892/0.891 \\ 
   GraphIQA   &0.849/0.850  &0.916/0.915& 0.888/0.881  
  \\  \midrule
    KSVQE &\textbf{0.867/.869} &\textbf{0.922/0.921} & \textbf{0.900/0.912}\\
    
    \bottomrule
  \end{tabular}}

\end{table}

%% file: sec/6_conlusion.tex
\section{Conclusion}

In this work, we take the first step to investigate S-UGC VQA from both subjective and objective studies. To address key challenges of kaleidoscope content and various processing flows in S-UGC videos, we build a large-scale kaleidoscopic short-form video database, named KVQ, which covers the primary creation modes, common content scenarios, as well as sophisticated video processing workflows. Moreover, we propose KSVQE based on the content-distortion understanding to identify quality-aware regions and perceive complex distortions. Experimental results reveal the efficacy of KSVQE. We hope to inspire future research for advancing VQA algorithms in S-UGC.

%% file: sec/X_suppl.tex

\section*{Appendix}
\noindent Section~\ref{sec:detailofprocessing} clarifies the details of our practical and sophisticated video processing workflows. And Section~\ref{sec:FeatureAnalysis} provides the analysis of the six feature distributions in our KVQ database. Section~\ref{sec:Human Study} includes the details about the test setup and data cleaning process. Section~\ref{sec:KSVQEDetails} compasses the details about region selection in the Quality-aware Region Selection (QRS) module and implementation details of our proposed KSVQE. In section~\ref{sec:More Experiment Results}, subsection~\ref{sec:More QRS}, subsection~\ref{sec:More CaM} and subsection~\ref{sec:More DaM} provide more ablation studies for the QRS, Content-adaptive Modulation (CaM) and Distortion-aware modulation (DaM), respectively.
 
\section{Details of Our Video Processing Workflows}
\label{sec:detailofprocessing}
In contrast to previous UGC databases~\cite{MDVQA, UGC-VIDEO}, which primarily focus on simulated compression artifacts, our proposed KVQ database is significantly different since its processing workflows are consistent with the practical applied workflows in the typical short-form video platform. Our processing workflow is composed of three cascaded parts, including video enhancement module $\phi_{e}(\cdot)$, pre-processing module $\phi_{p}(\cdot)$, and transcoding module $\phi_{t}(\cdot)$.

\noindent\paragraph{Video Enhancement Module $\phi_{e}(\cdot)$} is composed of three commonly used enhancement algorithms in short-form video platforms: De-Blur, De-Noise, and De-Artifact algorithm, where De-Blur aims to enhance the texture details of videos, and De-Noise is utilized to remove the structure/non-structure noises that are harmful to human perception. The De-Artifact algorithm is exploited to reduce other-form degradations, such as block artifacts. 

\begin{table*}[t]
   \centering
   \footnotesize
   \caption{Comparison of various dimensions among different UGC datasets.}
   \setlength{\tabcolsep}{0.1mm}{
   \begin{tabular}{c|c|c|c|c}
     \toprule
     UGC database & Video Sources  & Num Ref/Dis & Distortion Type & Subjective Form \\  
     \midrule
     CVD2014~\cite{CVD2014}& Captured & -/234  & authentic & MOS \\ 
     LIVE-VQC~\cite{LIVE-VQC}  &Captured & -/585 & authentic & MOS \\ 
     \midrule
     KoNViD-1k~\cite{KoNViD-1k}  & Flicker & -/1200  & authentic (UGC) & MOS\\ 
     YouTube-UGC~\cite{youtubeUGC}  & YouTube & -/1380  & authentic (UGC) & MOS\\ 
     Youku-V1K~\cite{Internet} & Youku& -/1072   & authentic (UGC) & MOS\\ 
     LSVQ~\cite{LSVQ} & IA,Flicker& -/39075   & authentic (UGC) & MOS\\  
     \midrule
     UGC-VIDEO~\cite{UGC-VIDEO} & TikTok& 50/550 & authentic+compression & MOS \\ 
     TaoLive~\cite{MDVQA} & Taobao& 418/3762  & authentic+compression & MOS \\ 
     KVQ & Short-form video platform & 600/3600 & authentic+enhancement+pre-processing+compression & MOS+Rank  \\
    
   \bottomrule
   \end{tabular}}
   \label{tab:data_c}
 \end{table*}
\begin{table*}[t]
  \centering
  \footnotesize
  \caption{Annotation criteria for subjective labeling scores from 1 to 5.}
  \begin{tabular}{p{3cm}|p{12cm}}
    \toprule
    Score & Annotation criteria \\ 
    \midrule
    1~Bad & The Visual information within video content becomes challenging or impossible to distinguish.  \\ 
    \midrule
    2~Poor & The primary video content remains distinguishable but exhibits pronounced noise, block artifacts, and blurriness, along with substantial jitter and lag. \\ 
    \midrule
    3~Fair & The primary video content is reasonably clear, but it includes noticeable distortions such as conspicuous noise, visual blurring, minor localized glare, or distinct edge sharpening. Additionally, the video exhibits a markedly blurry background texture. \\ 
    \midrule
    4~Good & The videos feature a clear primary subject, free from substantial noise or visual blurring, and devoid of apparent distortions such as jitter or glare. However, they exhibit limited overall textural complexity. \\ 
    \midrule
    5~Excellent & The primary video object is characterized by exceptional clarity, devoid of noise, block artifacts, blurriness, jitter, glare, or lag. It presents a high-quality spectacle distinguished by lucid textural elements. \\
    \bottomrule
  \end{tabular}
  \label{tab:label}
\end{table*}

\noindent\paragraph{Video Pre-processing Algorithms $\phi_{p}(\cdot)$} aims to reduce the high-frequency components that do not affect the human perception (\egno, the non-ROI region) or the high-frequency distortions, such as noises. In this way, it can reduce the compression and transmission costs while preserving/improving the subjective quality of short-form videos.  We select two pre-processing algorithms: global level pre-processing and region-of-interest (ROI) level pre-processing. The former is aimed at removing high-frequency information related to global-level impairment, while the latter focuses on eliminating high-frequency information associated with local-level impairment.

\noindent\paragraph{Video Transcoding Algorithms $\phi_{t}(\cdot)$}
The quantization parameters (QP) are the crucial parameters used to adjust the compression ratios, where higher QP corresponds to a higher compression ratio and lower visual quality. However, it is costly and labor-intensive to compress each video traversing each QP value (\ieno, from 0-51). To mitigate this and ensure the diversity of QP values, we divided the commonly-used QP range (\ieno, 16-47) into six intervals, encompassing 16-23, 24-31, 32-35, 36-39, 40-43, and 44-47, and then randomly select one QP from each interval for the compression of each video.

\begin{figure}
    \centering
    \includegraphics[width=0.5\textwidth]{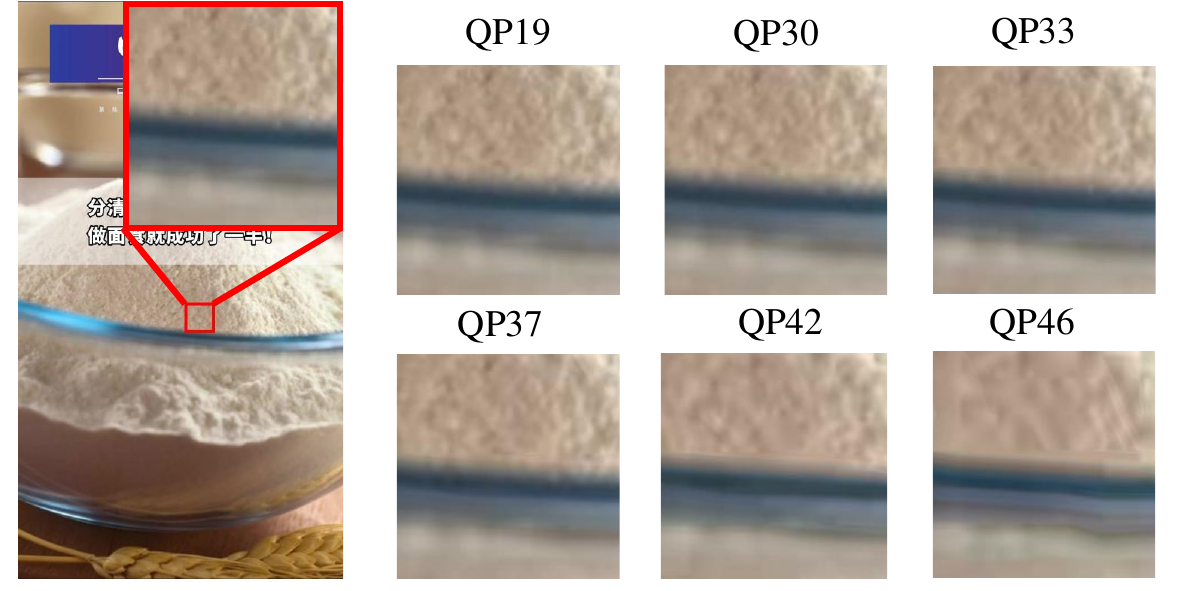}
    \caption{Examples of transcoding.}  
    \label{fig:C}
    \vspace{-3mm}
\end{figure}

\begin{figure}
    \centering
    \includegraphics[width=0.5\textwidth]{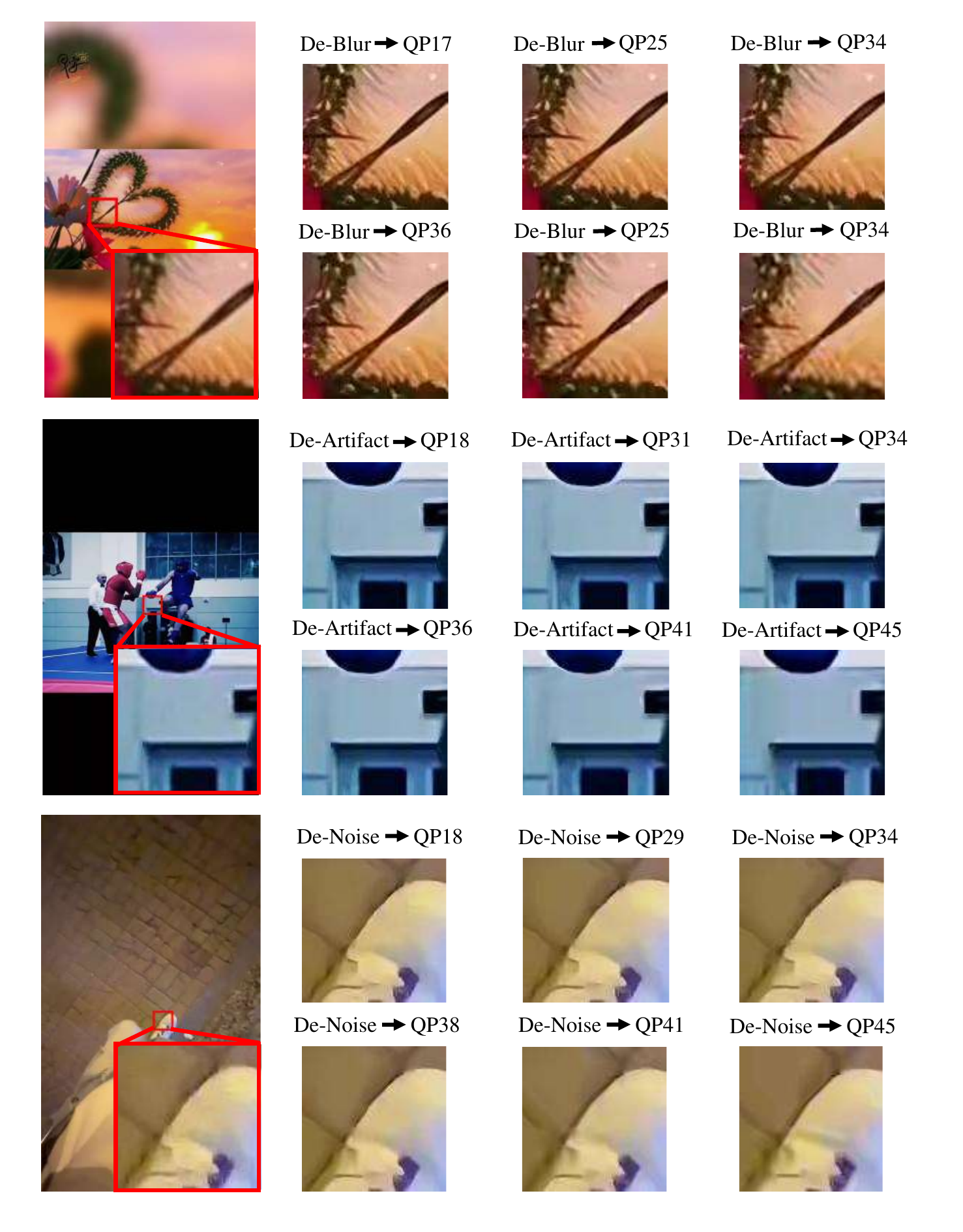}
    \caption{Examples of enhancement$\rightarrow$transcode.}  
    \label{fig:AC}
    \vspace{-3mm}
\end{figure}

\begin{figure*}
    \centering
    \includegraphics[width=1\textwidth]{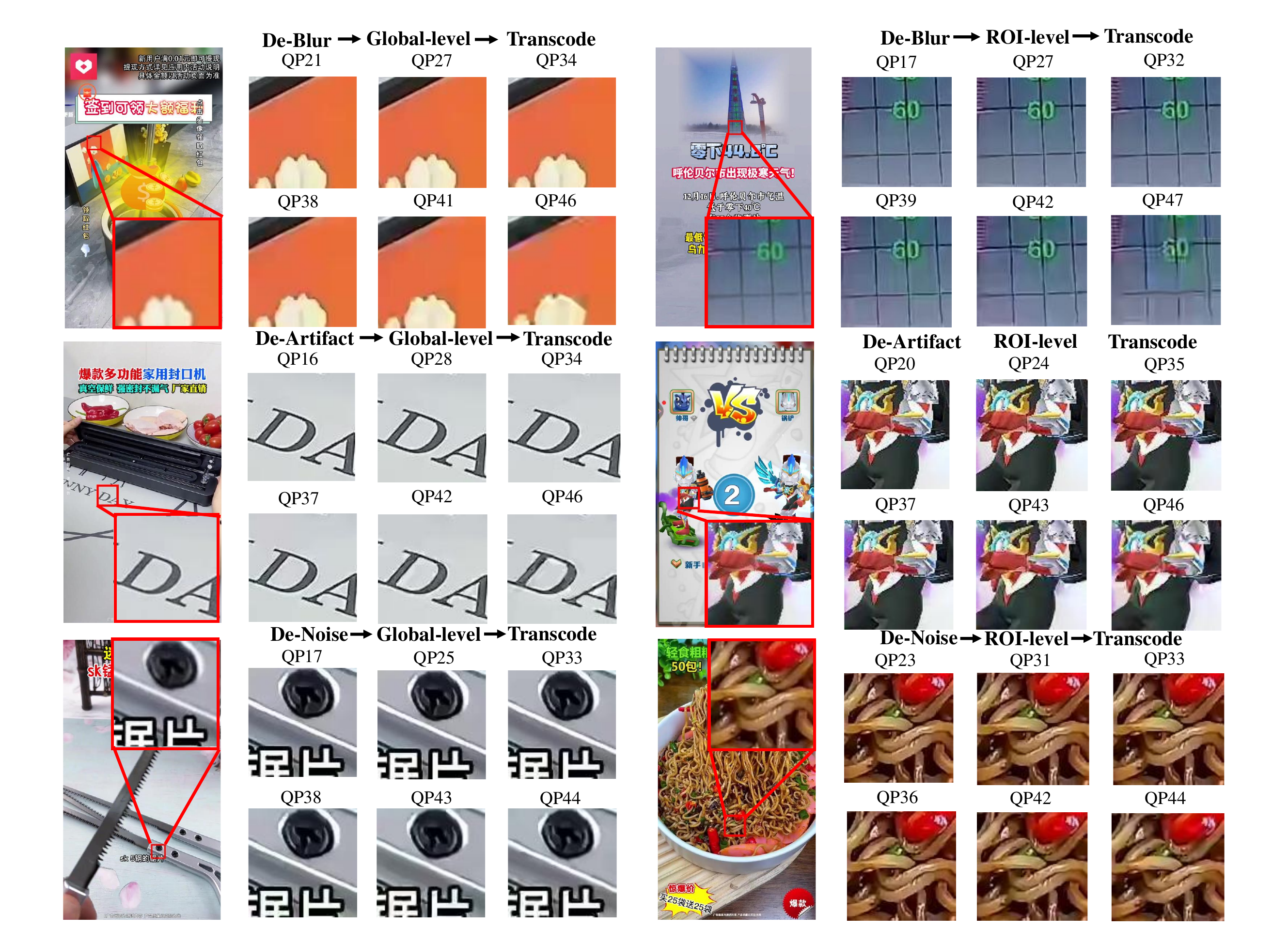}
    \caption{Examples of enhancement$\rightarrow$pre-processing$\rightarrow$transcode.}  
    \label{fig:ABC}
    \vspace{-3mm}
\end{figure*}

To demonstrate the effects of different processing workflows, we provide some examples for our three typical processing workflows in Fig. 2 of our manuscript, \ieno, $\phi_t(\cdot)$, $\phi_t(\phi_e(\cdot))$ and $\phi_t(\phi_p(\phi_e(\cdot)))$), Concretely, the example for $\phi_t(\cdot)$ is shown in Fig.~\ref{fig:C}. The example for $\phi_t(\phi_e(\cdot))$ is shown in Fig.~\ref{fig:AC}, and the example for $\phi_t(\phi_p(\phi_e(\cdot)))$ is shown in Fig.~\ref{fig:ABC}. The distorted patches are indicated by red boxes. Therefore, the KVQ dataset we established not only encompasses rich content within short video scenes but also spans more intricate video processing workflows, as illustrated in the comparisons across various UGC datasets in Table~\ref{tab:data_c}.

\section{Feature Analysis}
\label{sec:FeatureAnalysis}

\begin{figure}
    \centering
    \includegraphics[width=0.8\linewidth]{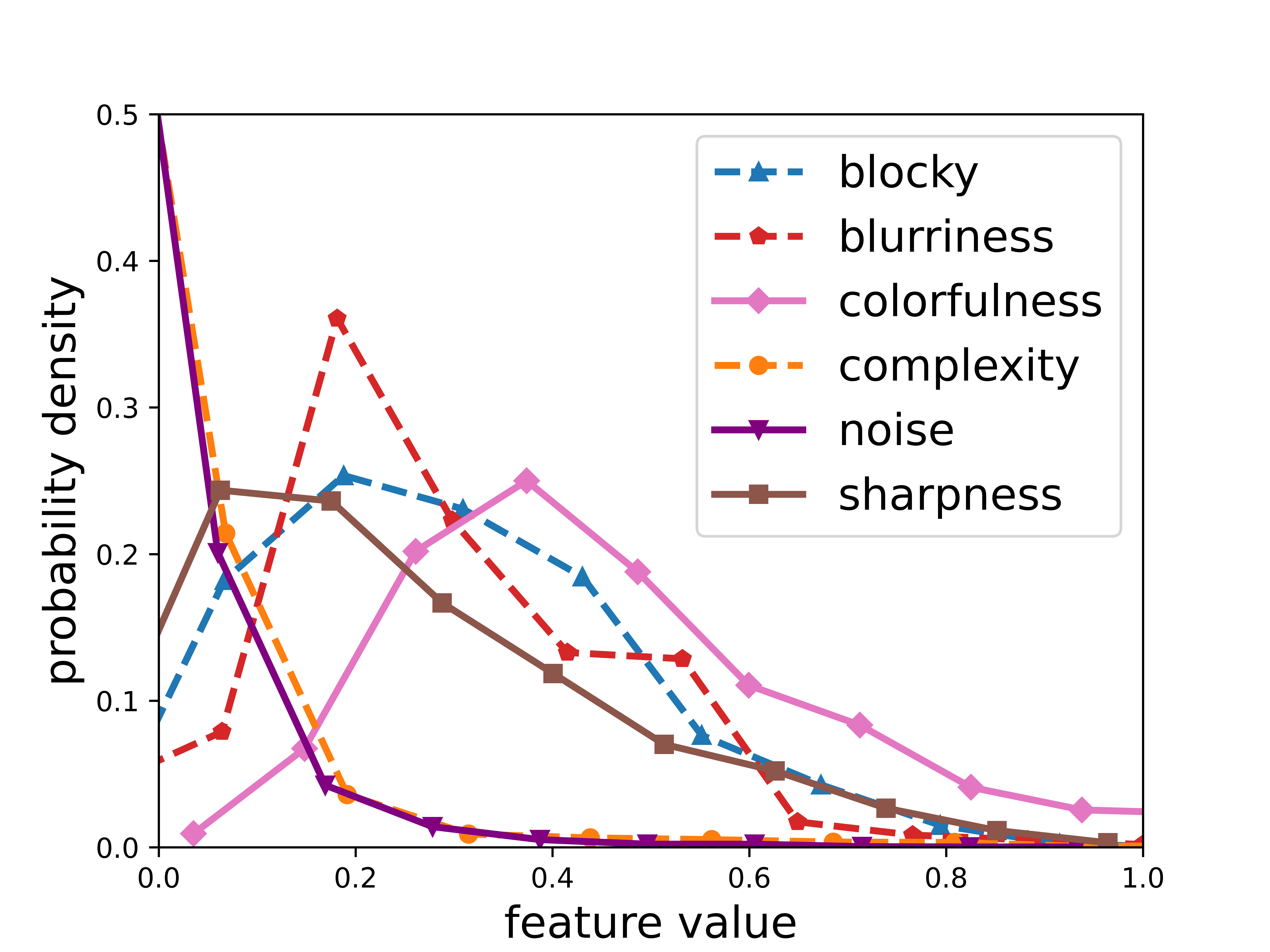}
    \caption{The feature distributions on the KVQ dataset.}
    \label{fig:6featurekvq}
    \vspace{-3mm}
\end{figure}

In summary, our KVQ database exhibits diverse feature characteristics across six video quality-related features, namely sharpness, blocky, blurriness, colorfulness, complexity, and noise. The distribution analysis, illustrated in Fig.~\ref{fig:6featurekvq}, highlights that the majority of features span a wide range, showcasing the feature diversity inherent in our database. Notably, blocky features and colorfulness features are more skewed towards the right, indicating a substantial presence of computer graphics, portraits, rich special effects, and common compression distortions, particularly on short-form video platforms. While complexity distribution and noise distribution skew towards lower values, the other features maintain closer adherence to middle values, with less pronounced spikes, providing an approximated overview of the distinctive feature characteristics on the typical short-form video platform.

 
\begin{figure}
    \centering
    \includegraphics[width=1\linewidth]{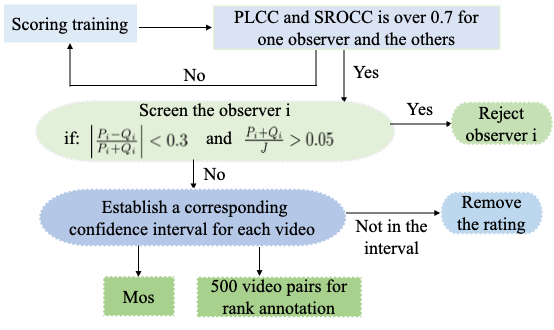}
    \caption{The overall data clean workflow. First, we ensure that the annotator achieves a correlation of 0.7, and then we screen the annotator via ITU-R BT 500.13 to confirm reliability. Finally, for each video, we set a corresponding confidence interval, scores that are outside this range will be removed.}
    \label{fig:scoring}
    \vspace{-3mm}
\end{figure}
\begin{figure*}
    \centering
    \includegraphics[width=0.7\textwidth]{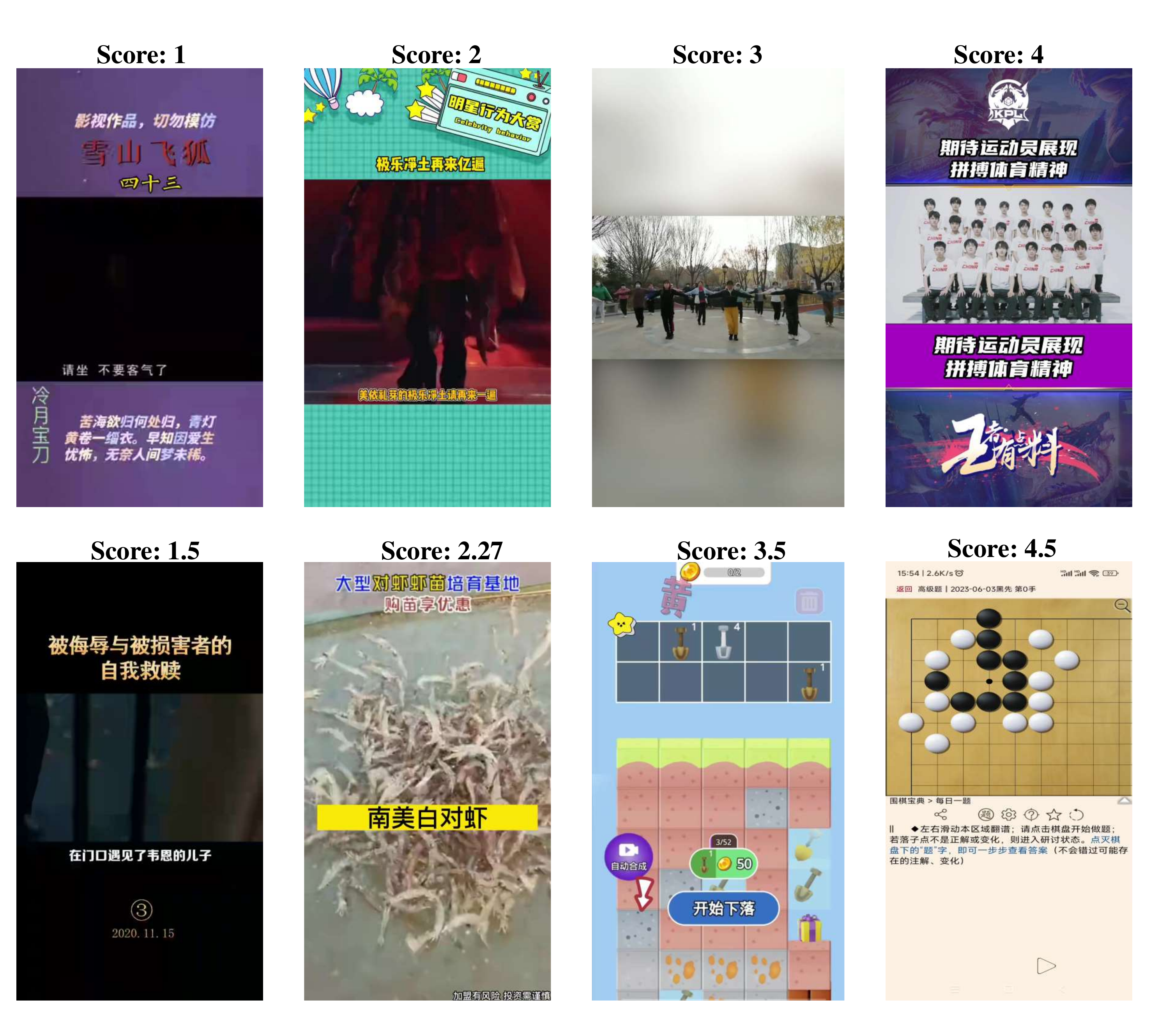}
    \caption{The video examples of quality score ranged from 1 to 5. }  
    \label{fig:framework}
    \vspace{-3mm}
\end{figure*}

\begin{figure*}
    \centering
    \includegraphics[width=0.7\textwidth]{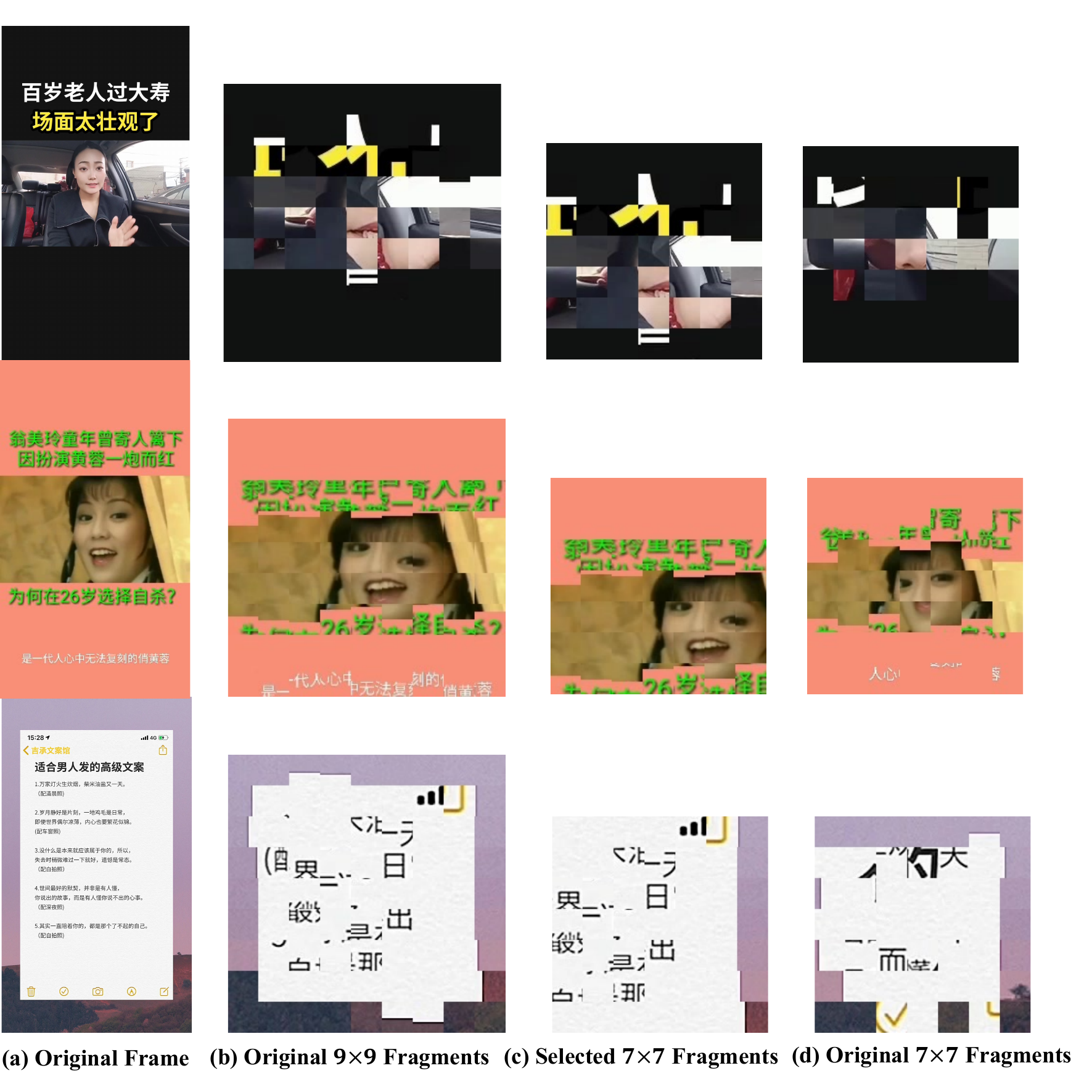}
    \caption{The visualization of original fragments of $7\times7$, $9 \times 9$ and the selected fragments of $7\times7$. }  
    \label{fig:QRS_v}
    \vspace{-3mm}
\end{figure*}
\section{More Detail About Human Study}
\label{sec:Human Study}
\subsection{Test Setup} 
The subjective experimentation involves a group of 15 observers, each tasked with assessing 4,200 videos. \tcb{Uniform MAC devices are employed by the observers to ensure standardized screen brightness and video resolution support.} During the scoring process, \tcb{a consistent stimulus evaluation method is utilized}, allowing for repeated viewing of the same video, ultimately leading to the evaluation of all videos. Continuous scales with intermediate numerical labels (ranging from 1 to 5 with a step size of 0.5) are employed for scoring.

Given the distinctive characteristics of the KVQ database, we establish the following instructions:
\begin{itemize}
    \item Scoring of special effects is lenient, disregarding the impact of special effects on perceptual quality, such as variations in lighting and shadows.
    \item In cases where a video is composed of multiple segments of merged content and significant quality differences, the total video score is computed based on the proportional duration of each segment.
    \item For videos containing text, the evaluation takes into account both text and background distortions, simultaneously determining whether the primary focus of the video is on text or background to derive the final quality score.
    \item For three-stage videos, only the content of the middle region is considered, with no regard for the influence of unrelated content in the upper and lower regions.
\end{itemize}

The guidelines for scoring, outlining the rules for assigning scores ranging from 1 to 5, are given in Table~\ref{tab:label}.

\subsection{Data Clean}
Fig.~\ref{fig:scoring} depicts the pipeline of the data cleaning.
First, it is necessary to ascertain the reliability of each observer's scores. After scoring overall videos in KVQ database, the correlation metrics (\ieno, SROCC, and PLCC) between an observer's scores and the average scores of other observers are computed. If the correlation falls below 0.7, retraining will be conducted for these observers.

Secondly, in accordance with ITU-R BT 500.13, further data processing is performed to screen observers for each video. Specifically, for each video, we compute the kurtosis of the scores to assess whether the ratings exhibit a normal distribution. Subsequently, based on the distribution of ratings, we calculate the quality score range for each video as 2 times the standard deviation from ratings or $\sqrt{20}$ times the standard deviation of the ratings. Based on this, We can determine the number of videos rated out of this range by each observer.
For the $i$-th observer:

 \noindent For all $j$ in $J$:
    
    If $u_{ij}\ge\overline{u}_{j}+\alpha*S_{j}$, then $P_{i}=P_{i}+1$. 
    
    If $u_{ij}\le\overline{u}_{j}+\alpha*S_{j}$, then $Q_{i}=Q_{i}+1$.

If $\left|\frac{P_{i}-Q_{i}}{P_{i}+Q_{i}}\right|<0.3$ and  
$\frac{P_{i}+Q_{i}}{J}>0.05$, the annotation made by the $i$-th observer will be rejected.

\noindent Here, $P_{i}$ denotes the number of videos that an observer has rated above the range, $Q_{i}$ represents the number rated below the range, and $J$ signifies the total number of videos rated by the observer, $\alpha$ can be 2 or $\sqrt{20}$, $S$ represents the standard deviation of each video. Following this step, we ascertain that all observers are reliable.

Thirdly, for each video in the KVQ dataset, it is imperative to establish a corresponding confidence interval for all ratings. This interval relies on the standard deviation and mean quality score of each video. We opt for a 95\% confidence interval, derived from: 
\begin{equation}
S_{j}=\sqrt{\sum_{i=1}^{N}\frac{(u_{ij}-\overline{u}_{j})^{2}}{(N-1)}}
\end{equation}

which yields the standard deviation. Subsequently, we calculate the range of the 95\% confidence interval as:
\begin{equation}
(\overline{u}_{j}-\delta_{j},\overline{u}_{j}+\delta_{j})
\end{equation}
with $\delta_{j}=1.96\frac{S_{j}}{\sqrt{N}}$, where $\overline{u}_{j}$ signifies the average rating for the $j^{th}$ video, $N$ is the number of observers that participate in the labeling of the $j^{th}$ video, $S_{j}$ represents the standard deviation of the $j^{th}$ video. The ratings falling outside the 95\% confidence interval range are then removed.

\section{Details of Our Framework KSVQE}
\label{sec:KSVQEDetails}
\subsection{QRS Details}
\paragraph{Spatial Region Selection}
The process involves selecting the most important fragments based on a quality-aware semantic importance score $I \in \mathbb{R}^{N}$, which contains two key points: i) how to make the selection operation differentiable, ii) how to preserve the spatial dependency within selected fragments.
To preserve the problem of spatial dependency, following work~\cite{stts}, we select the most quality-aware fragments through an aggregate-then-select strategy to simulate the Top-k selection of fragments with a correct spatial dependency. The aggregation operation involves splitting the reshaped score map $I \in \mathbb{R}^{\sqrt{N}\times \sqrt{N}}$ into a list of non-overlapping smaller score maps. Then average pooling is applied to each of these smaller score maps and obtains the patch importance vector $\hat{I}_r$. After the importance score aggregation operation, we apply the TopK operator to obtain the most quality-aware patches $\hat{X} $. We denote the TopK operator~\cite{cordonnier2021differentiable} as :
\begin{equation}
   inds=\mathrm{TopK}(\hat{I}_r)
\end{equation}
However, the $inds$ from the TopK operation are non-differentiable.

Inspired by the perturbed maximum method~\cite{DifferentiablePertubed}, the differentiable TopK can be realized through the solution for inputs with perturbation. 
The differentiable TopK operation shares a fundamental similarity with the Gumbel Softmax operation~\cite{jang2016categorical}. 
Specially, we sample uniform Gaussian noise $Z$ and add it to input $\hat{I}_r$,  then we can obtain the perturbed maximizer:
\begin{equation}
   Y_{inds,\sigma}= \mathop{\arg\max}\limits_{Y_{inds}\in \mathcal{C}} \langle Y_{inds}, \hat{I}_r+\sigma Z \rangle
\end{equation}
Where $Y_{inds}$ is the one-hot vector of indices $inds$ , $\sigma$ is a hyper-parameter to control the level of added noise. And we fix the $\sigma=0.5$ in our all experiments. And $\mathcal{C}$ is the convex polytope constrain set.
For backward, the gradient can be passed from variable parameter $Z$ to optimization variable $\hat{I}_r$. 
\subsection{Implementation Details}

The semantic adapter $f$ and the distortion adapter $f_d$ each consist of several fully-connected (FC) layers with dimensional variations of '768-192-768' and '128-32-768', respectively. The Multi-Head Cross Attention for both semantic modulation and distortion modulation has a head number of 8 and a dimension of 768. The Multi-Head Self Attention in distortion modulation has a head number of 8 and a dimension of 768. The modulation parameter generator $l_{ss}$ and $l_{so}$ for semantic modulation are the convolutions with the kernel size of $1\times 1$ and dimension variation of ``768-1". And the modulation parameter generator $l_{ds}$ and $l_{do}$ for distortion modulation are the FC layer with the dimension variation of ``768-768".

For $N_t$ key frames in the input of the CLIP visual encoder, we partition the videos into segments and select a single frame from each segment to encapsulate the comprehensive semantic information of that segment. Subsequently, utilizing the acquired quality-aware importance vector and visual tokens for modulation guidance, we extend the temporal dimension of $N_t$ to $T$.

In the training process, we utilize AdamW optimizer~\cite{loshchilov2018fixing} with a learning rate of $3\times e^{-5}$ and a weight decay of $0.05$ for optimization. And batchsize set as 8.

\section{More Experiment Results}
\label{sec:More Experiment Results}
\subsection{QRS}
\label{sec:More QRS}
\paragraph{ More Variants About QRS}

In order to investigate the optimal original number of fragments for region selection in QRS, we compare multiple numbers of original fragments in Table~\ref{tab:QRS_num}. Notably, we observe that extracting $7 \times 7$ fragments from $9 \times 9$ input fragments for the 3D Swin Transformer yielded the most optimal performance. 
In the context of region selection, selecting an excessive number of original fragments results in the retention of excessive redundant information. Conversely, opting for too few numbers of original fragments leads to the absence of crucial, quality-aware visual information. The selection of $9 \times 9$ for original fragments strikes a balance, demonstrating superior performance by capturing essential visual features without succumbing to information redundancy or loss of significance.
\begin{table}[]
\setlength{\tabcolsep}{0.3mm}{
\caption{Ablation study for the number of fragments in QRS.}
\vspace{-0.3cm}
\label{tab:QRS_num}
\footnotesize
 \begin{center}
\begin{tabular}{c|cc|cc|cc}
\toprule
\multicolumn{1}{c|}{\multirow{2}{*}{Region Selection}} & \multicolumn{2}{c|}{KVQ}         & \multicolumn{2}{c|}{KoNViD-1k}       & \multicolumn{2}{c}{YouTube-UGC}    \\ 
\multicolumn{1}{c|}{}                                 & \multicolumn{1}{c}{SROCC} & PLCC & \multicolumn{1}{c}{SROCC} & PLCC & \multicolumn{1}{c}{SROCC} & PLCC \\ \midrule
$8 \times 8 \rightarrow 7\times 7$                                                  & \multicolumn{1}{c}{0.841}      & 0.841     & \multicolumn{1}{c}{\textbf{0.918}}      & 0.917     & \multicolumn{1}{c}{0.887}     & 0.903    \\
$9 \times 9 \rightarrow 7 \times 7$                                 &\multicolumn{1}{c}{\textbf{0.847}}      &\textbf{0.853}      & \multicolumn{1}{c}{0.917}      & \textbf{0.920}     & \multicolumn{1}{c}{\textbf{0.894}}      & \textbf{0.906}    \\ 
$10 \times 10 \rightarrow 7 \times 7$                                            & \multicolumn{1}{c}{0.847}     &0.848      & \multicolumn{1}{c}{0.911 }      & 0.914   & \multicolumn{1}{c}{0.892}      & 0.907    \\ \bottomrule
\end{tabular} \end{center}}
\vspace{-3mm}
\end{table}

\paragraph{Visualization About QRS}
Also, we visualize the selected fragments and original fragments in Fig.~\ref{fig:QRS_v}. And we can see that for the first row of three-stage video, our Quality Region Selection (QRS) method excels in extracting concentrated regions within the central regions. In videos characterized by extensive monochromatic backgrounds (\ieno, the second row), QRS is capable of capturing visually enriched regions for the face and text. Additionally, when dealing with videos incorporating much text (\ieno, the third row), QRS focuses mainly on the text area rather than the background, which is consistent with human attention.

\subsection{CaM}
\label{sec:More CaM}
\paragraph{The effectiveness of adapter-style training}
To verify the effectiveness of our adapter on cls token, we conduct an experiment by removing it and comparing the results with those obtained using the full modules for content understanding, as illustrated in Table~\ref{tab:adapter_CaM}. The results show that adding adapter-style training can bring a performance gain of 0.007/0.004 on SROCC and PLCC on the KVQ database.  It illustrates that the feature adaptation to quality-aware space is necessary to incorporate content understanding and extract the quality-aware semantics to provide guidance.
\begin{table}[]
\setlength{\tabcolsep}{0.3mm}{
\caption{Ablation study for adapter-style training in CaM.}
\label{tab:adapter_CaM}
\footnotesize
 \begin{center}
\begin{tabular}{c|cc|cc|cc}
\toprule
\multicolumn{1}{c|}{\multirow{2}{*}{Content Adapter}} & \multicolumn{2}{c|}{KVQ}         & \multicolumn{2}{c|}{KoNViD-1k}       & \multicolumn{2}{c}{YouTube-UGC}    \\ 
\multicolumn{1}{c|}{}                                 & \multicolumn{1}{c}{SROCC} & PLCC & \multicolumn{1}{c}{SROCC} & PLCC & \multicolumn{1}{c}{SROCC} & PLCC \\ \midrule
QRS+CaM (w.o. adapter)                                                  & \multicolumn{1}{c}{0.841}      &0.850      & \multicolumn{1}{c}{ 0.913}      &0.914      & \multicolumn{1}{c}{0.882}     &0.893     \\
QRS+CaM                    
& \multicolumn{1}{c}{\textbf{0.848}}      &
\textbf{0.854}    & \multicolumn{1}{c}{\textbf{0.918}}      &\textbf{0.922}     & \multicolumn{1}{c}{ \textbf{0.895}}      & \textbf{0.901}     \\ 
 \bottomrule
\end{tabular} \end{center}}
\vspace{-3mm}
\end{table}
\begin{table}[]
\setlength{\tabcolsep}{0.3mm}{
\caption{Ablation study for multiple variants of selection in DaM.}

\label{tab:more_DaM}
\footnotesize
 \begin{center}
\begin{tabular}{c|cc|cc|cc}
\toprule
\multicolumn{1}{c|}{\multirow{2}{*}{DaM}} & \multicolumn{2}{c|}{KVQ}         & \multicolumn{2}{c|}{KoNViD-1k}       & \multicolumn{2}{c}{YouTube-UGC}    \\ 
\multicolumn{1}{c|}{}                                 & \multicolumn{1}{c}{SROCC} & PLCC & \multicolumn{1}{c}{SROCC} & PLCC & \multicolumn{1}{c}{SROCC} & PLCC \\ \midrule
CA+CM                                                  & \multicolumn{1}{c}{0.832}      &0.834      & \multicolumn{1}{c}{ 0.911}      &0.912      & \multicolumn{1}{c}{0.888}     &0.899     \\
CASA+CM (DaM)                                &\multicolumn{1}{c}{\textbf{0.839}}      &  \textbf{0.843}   & \multicolumn{1}{c}{\textbf{0.915}  }      &\textbf{0.914}  & \multicolumn{1}{c}{\textbf{0.893}}      & \textbf{0.910}    \\ 
 \bottomrule
\end{tabular} \end{center}}
\vspace{-0.3cm}
\end{table}

\subsection{DaM}
\label{sec:More DaM}
\paragraph{More Variants About DaM}
Also, we analyze the effectiveness of our distortion adapter on DaM in Table~\ref{tab:adapter_DaM}. We can see that adapter-style training demonstrates an improvement in the performance of 0.008/0.012 in terms of SROCC and PLCC on KVQ database. It reveals the significance of adapting knowledge from CONTRIQUE to distortion distribution in KVQ database.

For more variants for distortion modulation in DaM, we remove the multi-head self-attention as the variant ``CA+CM" and compare it with our DaM (\ieno, CASA+CM) in Table~\ref{tab:more_DaM}. The results show that DaM benefits from the influence exerted by self-attention for temporal distortion extraction, resulting in an augmented performance outcome of 0.007/0.009 in terms of SROCC and PLCC on KVQ database.

\begin{table}[]
\setlength{\tabcolsep}{0.3mm}{
\caption{Ablation study for adapter-style training in DaM.}
\vspace{-0.3cm}
\label{tab:adapter_DaM}
\footnotesize
 \begin{center}
\begin{tabular}{c|cc}
\toprule
\multicolumn{1}{c|}{\multirow{2}{*}{Distortion Adapter}} & \multicolumn{2}{c}{KVQ}         \\ 
\multicolumn{1}{c|}{}                                 & \multicolumn{1}{c}{SROCC} & PLCC   \\ \midrule
DaM (w.o. adapter)                                                  & \multicolumn{1}{c}{0.831}      &0.831      \\
DaM                               &\multicolumn{1}{c}{\textbf{0.839}}      &  \textbf{0.843}    \\ 
 \bottomrule
\end{tabular} \end{center}}

\end{table}

\subsection{The Combination of Content-Distortion Understanding}

In this section, we investigate another method to incorporate content prior and distortion prior into the original feature. 
We compared our proposed modulation method with the simplest fusion approach, concatenation, and the results are presented in Table~\ref{tab:concat}. The study results indicate that our modulation method is more effective in explicitly modeling the understanding of content and distortion. Consequently, it leads to an improvement in the performance of all databases.
\begin{table}[]
\setlength{\tabcolsep}{0.3mm}{
\caption{Ablation study for multiple variants of selection in combination.}
\vspace{-0.3cm}
\label{tab:concat}
\footnotesize
 \begin{center}
\begin{tabular}{c|cc|cc|cc}
\toprule
\multicolumn{1}{c|}{\multirow{2}{*}{DaM}} & \multicolumn{2}{c|}{KVQ}         & \multicolumn{2}{c|}{KoNViD-1k}       & \multicolumn{2}{c}{YouTube-UGC}    \\ 
\multicolumn{1}{c|}{}                                 & \multicolumn{1}{c}{SROCC} & PLCC & \multicolumn{1}{c}{SROCC} & PLCC & \multicolumn{1}{c}{SROCC} & PLCC \\ \midrule
QRS+concat                                                  & \multicolumn{1}{c}{0.853}      & 0.856     & \multicolumn{1}{c}{ 0.912}      &  0.914    & \multicolumn{1}{c}{0.891}     &0.894     \\
QRS+CaM+DaM                                 &\multicolumn{1}{c}{\textbf{0.867}}      &\textbf{0.869}    & \multicolumn{1}{c}{\textbf{ 0.922} }      & \textbf{0.921}  & \multicolumn{1}{c}{\textbf{0.900}}      &\textbf{0.912  }   \\  \bottomrule
\end{tabular} \end{center}}

\end{table}

%% file: main.bbl
\begin{thebibliography}{79}
\providecommand{\natexlab}[1]{#1}
\providecommand{\url}[1]{\texttt{#1}}
\expandafter\ifx\csname urlstyle\endcsname\relax
  \providecommand{\doi}[1]{doi: #1}\else
  \providecommand{\doi}{doi: \begingroup \urlstyle{rm}\Url}\fi

\bibitem[Agarwal et~al.(2021)Agarwal, Krueger, Clark, Radford, Kim, and Brundage]{CLIPsemantic}
Sandhini Agarwal, Gretchen Krueger, Jack Clark, Alec Radford, Jong~Wook Kim, and Miles Brundage.
\newblock Evaluating {CLIP:} towards characterization of broader capabilities and downstream implications.
\newblock \emph{CoRR}, abs/2108.02818, 2021.

\bibitem[Berthet et~al.(2020)Berthet, Blondel, Teboul, Cuturi, Vert, and Bach]{DifferentiablePertubed}
Quentin Berthet, Mathieu Blondel, Olivier Teboul, Marco Cuturi, Jean{-}Philippe Vert, and Francis~R. Bach.
\newblock Learning with differentiable pertubed optimizers.
\newblock In \emph{Advances in Neural Information Processing Systems 33: Annual Conference on Neural Information Processing Systems 2020, NeurIPS 2020, December 6-12, 2020, virtual}, 2020.

\bibitem[BT(2002)]{bt2002methodology}
RIR BT.
\newblock Methodology for the subjective assessment of the quality of television pictures.
\newblock \emph{International Telecommunication Union}, 4, 2002.

\bibitem[Chan et~al.(2022)Chan, Zhou, Xu, and Loy]{enhancement1}
Kelvin C.~K. Chan, Shangchen Zhou, Xiangyu Xu, and Chen~Change Loy.
\newblock Basicvsr++: Improving video super-resolution with enhanced propagation and alignment.
\newblock In \emph{{IEEE/CVF} Conference on Computer Vision and Pattern Recognition, {CVPR} 2022, New Orleans, LA, USA, June 18-24, 2022}, pages 5962--5971. {IEEE}, 2022.

\bibitem[Chen et~al.(2022{\natexlab{a}})Chen, Zhu, Li, Lu, Fan, and Wang]{GSTVQA}
Baoliang Chen, Lingyu Zhu, Guo Li, Fangbo Lu, Hongfei Fan, and Shiqi Wang.
\newblock Learning generalized spatial-temporal deep feature representation for no-reference video quality assessment.
\newblock \emph{{IEEE} Trans. Circuits Syst. Video Technol.}, 32\penalty0 (4):\penalty0 1903--1916, 2022{\natexlab{a}}.

\bibitem[Chen et~al.(2022{\natexlab{b}})Chen, Tao, Zhang, Wang, Ye, Wang, Hu, and Savvides]{Convadapter}
Hao Chen, Ran Tao, Han Zhang, Yidong Wang, Wei Ye, Jindong Wang, Guosheng Hu, and Marios Savvides.
\newblock Conv-adapter: Exploring parameter efficient transfer learning for convnets.
\newblock \emph{CoRR}, abs/2208.07463, 2022{\natexlab{b}}.

\bibitem[Chen et~al.(2022{\natexlab{c}})Chen, Li, Wu, Dong, and Shi]{contraiVideo}
Pengfei Chen, Leida Li, Jinjian Wu, Weisheng Dong, and Guangming Shi.
\newblock Contrastive self-supervised pre-training for video quality assessment.
\newblock \emph{{IEEE} Trans. Image Process.}, 31:\penalty0 458--471, 2022{\natexlab{c}}.

\bibitem[Cordonnier et~al.(2021)Cordonnier, Mahendran, Dosovitskiy, Weissenborn, Uszkoreit, and Unterthiner]{cordonnier2021differentiable}
Jean{-}Baptiste Cordonnier, Aravindh Mahendran, Alexey Dosovitskiy, Dirk Weissenborn, Jakob Uszkoreit, and Thomas Unterthiner.
\newblock Differentiable patch selection for image recognition.
\newblock In \emph{{CVPR}}, pages 2351--2360. Computer Vision Foundation / {IEEE}, 2021.

\bibitem[Fang et~al.(2020)Fang, Zhu, Zeng, Ma, and Wang]{SPAQ}
Yuming Fang, Hanwei Zhu, Yan Zeng, Kede Ma, and Zhou Wang.
\newblock Perceptual quality assessment of smartphone photography.
\newblock In \emph{2020 {IEEE/CVF} Conference on Computer Vision and Pattern Recognition, {CVPR} 2020, Seattle, WA, USA, June 13-19, 2020}, pages 3674--3683. Computer Vision Foundation / {IEEE}, 2020.

\bibitem[Gesmundo and Dean(2022)]{muNet}
Andrea Gesmundo and Jeff Dean.
\newblock munet: Evolving pretrained deep neural networks into scalable auto-tuning multitask systems.
\newblock \emph{CoRR}, abs/2205.10937, 2022.

\bibitem[Ghadiyaram et~al.(2018)Ghadiyaram, Pan, Bovik, Moorthy, Panda, and Yang]{LIVE-Qualcomm}
Deepti Ghadiyaram, Janice Pan, Alan~C. Bovik, Anush~Krishna Moorthy, Prasanjit Panda, and Kai{-}Chieh Yang.
\newblock In-capture mobile video distortions: {A} study of subjective behavior and objective algorithms.
\newblock \emph{{IEEE} Trans. Circuits Syst. Video Technol.}, 28\penalty0 (9):\penalty0 2061--2077, 2018.

\bibitem[Guo et~al.(2023)Guo, Feng, Zhang, Jin, and Chen]{RLVC}
Zongyu Guo, Runsen Feng, Zhizheng Zhang, Xin Jin, and Zhibo Chen.
\newblock Learning cross-scale weighted prediction for efficient neural video compression.
\newblock \emph{{IEEE} Trans. Image Process.}, 32:\penalty0 3567--3579, 2023.

\bibitem[Hosu et~al.(2017)Hosu, Hahn, Jenadeleh, Lin, Men, Szir{\'{a}}nyi, Li, and Saupe]{KoNViD-1k}
Vlad Hosu, Franz Hahn, Mohsen Jenadeleh, Hanhe Lin, Hui Men, Tam{\'{a}}s Szir{\'{a}}nyi, Shujun Li, and Dietmar Saupe.
\newblock The konstanz natural video database (konvid-1k).
\newblock In \emph{QoMEX}, pages 1--6. {IEEE}, 2017.

\bibitem[Installations and Line(1999)]{installations1999subjective}
Telephone Installations and Local Line.
\newblock Subjective video quality assessment methods for multimedia applications.
\newblock \emph{Networks}, 910\penalty0 (37):\penalty0 5, 1999.

\bibitem[Jang et~al.(2016)Jang, Gu, and Poole]{jang2016categorical}
Eric Jang, Shixiang Gu, and Ben Poole.
\newblock Categorical reparameterization with gumbel-softmax.
\newblock \emph{arXiv preprint arXiv:1611.01144}, 2016.

\bibitem[Ji et~al.(2019)Ji, Wu, Shi, Wan, and Xie]{SemanticIQA}
Weiping Ji, Jinjian Wu, Guangming Shi, Wenfei Wan, and Xuemei Xie.
\newblock Blind image quality assessment with semantic information.
\newblock \emph{J. Vis. Commun. Image Represent.}, 58:\penalty0 195--204, 2019.

\bibitem[Jiang et~al.(2023)Jiang, Sang, Hu, and Liu]{Self_Supervised}
Shaojie Jiang, Qingbing Sang, Zongyao Hu, and Lixiong Liu.
\newblock Self-supervised representation learning for video quality assessment.
\newblock \emph{{IEEE} Trans. Broadcast.}, 69\penalty0 (1):\penalty0 118--129, 2023.

\bibitem[Kancharla and Channappayya(2022)]{STEM}
Parimala Kancharla and Sumohana~S. Channappayya.
\newblock Completely blind quality assessment of user generated video content.
\newblock \emph{{IEEE} Trans. Image Process.}, 31:\penalty0 263--274, 2022.

\bibitem[Korhonen(2019)]{TLVQM}
Jari Korhonen.
\newblock Two-level approach for no-reference consumer video quality assessment.
\newblock \emph{{IEEE} Trans. Image Process.}, 28\penalty0 (12):\penalty0 5923--5938, 2019.

\bibitem[Li et~al.(2022)Li, Zhang, Tian, Zhai, and Wang]{BVQA}
Bowen Li, Weixia Zhang, Meng Tian, Guangtao Zhai, and Xianpei Wang.
\newblock Blindly assess quality of in-the-wild videos via quality-aware pre-training and motion perception.
\newblock \emph{{IEEE} Trans. Circuits Syst. Video Technol.}, 32\penalty0 (9):\penalty0 5944--5958, 2022.

\bibitem[Li et~al.(2019{\natexlab{a}})Li, Jiang, and Jiang]{VSFA}
Dingquan Li, Tingting Jiang, and Ming Jiang.
\newblock Quality assessment of in-the-wild videos.
\newblock In \emph{{ACM} Multimedia}, pages 2351--2359. {ACM}, 2019{\natexlab{a}}.

\bibitem[Li et~al.(2019{\natexlab{b}})Li, Jiang, Lin, and Jiang]{BlueSky}
Dingquan Li, Tingting Jiang, Weisi Lin, and Ming Jiang.
\newblock Which has better visual quality: The clear blue sky or a blurry animal?
\newblock \emph{{IEEE} Trans. Multim.}, 21\penalty0 (5):\penalty0 1221--1234, 2019{\natexlab{b}}.

\bibitem[Li et~al.(2021)Li, Jiang, and Jiang]{unfiedVQA}
Dingquan Li, Tingting Jiang, and Ming Jiang.
\newblock Unified quality assessment of in-the-wild videos with mixed datasets training.
\newblock \emph{Int. J. Comput. Vis.}, 129\penalty0 (4):\penalty0 1238--1257, 2021.

\bibitem[Li et~al.(2023{\natexlab{a}})Li, Lian, Lu, Bai, Chen, and Wang]{li2023graphadapter}
Xin Li, Dongze Lian, Zhihe Lu, Jiawang Bai, Zhibo Chen, and Xinchao Wang.
\newblock Graphadapter: Tuning vision-language models with dual knowledge graph.
\newblock \emph{arXiv preprint arXiv:2309.13625}, 2023{\natexlab{a}}.

\bibitem[Li et~al.(2023{\natexlab{b}})Li, Lu, and Chen]{li2023freqalign}
Xin Li, Yiting Lu, and Zhibo Chen.
\newblock Freqalign: Excavating perception-oriented transferability for blind image quality assessment from a frequency perspective.
\newblock \emph{IEEE Transactions on Multimedia}, 2023{\natexlab{b}}.

\bibitem[Li et~al.(2020)Li, Meng, Zhang, Wang, Wang, and Ma]{UGC-VIDEO}
Yang Li, Shengbin Meng, Xinfeng Zhang, Shiqi Wang, Yue Wang, and Siwei Ma.
\newblock {UGC-VIDEO:} perceptual quality assessment of user-generated videos.
\newblock In \emph{3rd {IEEE} Conference on Multimedia Information Processing and Retrieval, {MIPR} 2020, Shenzhen, China, August 6-8, 2020}, pages 35--38. {IEEE}, 2020.

\bibitem[Liang et~al.(2023)Liang, Wu, Dai, Li, Zhao, Zhang, Zhang, Vajda, and Marculescu]{CLIPsemantic1}
Feng Liang, Bichen Wu, Xiaoliang Dai, Kunpeng Li, Yinan Zhao, Hang Zhang, Peizhao Zhang, Peter Vajda, and Diana Marculescu.
\newblock Open-vocabulary semantic segmentation with mask-adapted {CLIP}.
\newblock In \emph{{CVPR}}, pages 7061--7070. {IEEE}, 2023.

\bibitem[Liao et~al.(2022)Liao, Xu, Wu, Chen, Sun, Yan, and Lin]{tpqi}
Liang Liao, Kangmin Xu, Haoning Wu, Chaofeng Chen, Wenxiu Sun, Qiong Yan, and Weisi Lin.
\newblock Exploring the effectiveness of video perceptual representation in blind video quality assessment.
\newblock In \emph{{MM} '22: The 30th {ACM} International Conference on Multimedia, Lisboa, Portugal, October 10 - 14, 2022}, pages 837--846. {ACM}, 2022.

\bibitem[Ling et~al.(2020)Ling, Baveye, Callet, Skinner, and Katsavounidis]{transcode1}
Suiyi Ling, Yoann Baveye, Patrick~Le Callet, Jim Skinner, and Ioannis Katsavounidis.
\newblock Towards perceptually-optimized compression of user generated content {(UGC):} prediction of {UGC} rate-distortion category.
\newblock In \emph{{IEEE} International Conference on Multimedia and Expo, {ICME} 2020, London, UK, July 6-10, 2020}, pages 1--6. {IEEE}, 2020.

\bibitem[Liu et~al.(2023)Liu, Wu, Yuan, Sun, Tang, Zheng, Wen, and Li]{ADAQA}
Hongbo Liu, Mingda Wu, Kun Yuan, Ming Sun, Yansong Tang, Chuanchuan Zheng, Xing Wen, and Xiu Li.
\newblock Ada-dqa: Adaptive diverse quality-aware feature acquisition for video quality assessment.
\newblock \emph{CoRR}, abs/2308.00729, 2023.

\bibitem[Liu et~al.(2022)Liu, Li, An, and Chen]{liu2022sourceBIQA}
Jianzhao Liu, Xin Li, Shukun An, and Zhibo Chen.
\newblock Source-free unsupervised domain adaptation for blind image quality assessment.
\newblock \emph{arXiv preprint arXiv:2207.08124}, 2022.

\bibitem[Liu et~al.(2018)Liu, Duanmu, and Wang]{compressedVQA}
Wentao Liu, Zhengfang Duanmu, and Zhou Wang.
\newblock End-to-end blind quality assessment of compressed videos using deep neural networks.
\newblock In \emph{2018 {ACM} Multimedia Conference on Multimedia Conference, {MM} 2018, Seoul, Republic of Korea, October 22-26, 2018}, pages 546--554. {ACM}, 2018.

\bibitem[Liu et~al.(2021)Liu, Liu, Gu, Zhang, Wu, Qiao, and Dong]{liu2021discovering}
Yihao Liu, Anran Liu, Jinjin Gu, Zhipeng Zhang, Wenhao Wu, Yu Qiao, and Chao Dong.
\newblock Discovering distinctive" semantics" in super-resolution networks.
\newblock \emph{arXiv preprint arXiv:2108.00406}, 2021.

\bibitem[Loshchilov and Hutter(2018)]{loshchilov2018fixing}
Ilya Loshchilov and Frank Hutter.
\newblock Fixing weight decay regularization in adam.
\newblock 2018.

\bibitem[Madhusudana et~al.(2022)Madhusudana, Birkbeck, Wang, Adsumilli, and Bovik]{CONTRIQUE}
Pavan~C. Madhusudana, Neil Birkbeck, Yilin Wang, Balu Adsumilli, and Alan~C. Bovik.
\newblock Image quality assessment using contrastive learning.
\newblock \emph{{IEEE} Trans. Image Process.}, 31:\penalty0 4149--4161, 2022.

\bibitem[Mittal et~al.(2016)Mittal, Saad, and Bovik]{VIIDEO}
Anish Mittal, Michele~A. Saad, and Alan~C. Bovik.
\newblock A completely blind video integrity oracle.
\newblock \emph{{IEEE} Trans. Image Process.}, 25\penalty0 (1):\penalty0 289--300, 2016.

\bibitem[Nuutinen et~al.(2016)Nuutinen, Virtanen, Vaahteranoksa, Vuori, Oittinen, and H{\"{a}}kkinen]{CVD2014}
Mikko Nuutinen, Toni Virtanen, Mikko Vaahteranoksa, Tero Vuori, Pirkko Oittinen, and Jukka H{\"{a}}kkinen.
\newblock {CVD2014} - {A} database for evaluating no-reference video quality assessment algorithms.
\newblock \emph{{IEEE} Trans. Image Process.}, 25\penalty0 (7):\penalty0 3073--3086, 2016.

\bibitem[Pavez et~al.(2022)Pavez, Perez, Xiong, Ortega, and Adsumilli]{DBLP:conf/icip/PavezPXOA22}
Eduardo Pavez, Enrique Perez, Xin Xiong, Antonio Ortega, and Balu Adsumilli.
\newblock Compression of user generated content using denoised references.
\newblock In \emph{{ICIP}}, pages 4188--4192. {IEEE}, 2022.

\bibitem[Radford et~al.(2021)Radford, Kim, Hallacy, Ramesh, Goh, Agarwal, Sastry, Askell, Mishkin, Clark, Krueger, and Sutskever]{CLIP}
Alec Radford, Jong~Wook Kim, Chris Hallacy, Aditya Ramesh, Gabriel Goh, Sandhini Agarwal, Girish Sastry, Amanda Askell, Pamela Mishkin, Jack Clark, Gretchen Krueger, and Ilya Sutskever.
\newblock Learning transferable visual models from natural language supervision.
\newblock In \emph{{ICML}}, pages 8748--8763. {PMLR}, 2021.

\bibitem[Rasheed et~al.(2022)Rasheed, Maaz, Khattak, Khan, and Khan]{CLIPsemantic2}
Hanoona~Abdul Rasheed, Muhammad Maaz, Muhammad~Uzair Khattak, Salman~H. Khan, and Fahad~Shahbaz Khan.
\newblock Bridging the gap between object and image-level representations for open-vocabulary detection.
\newblock In \emph{NeurIPS}, 2022.

\bibitem[Saad et~al.(2014)Saad, Bovik, and Charrier]{V-BLINDS}
Michele~A. Saad, Alan~C. Bovik, and Christophe Charrier.
\newblock Blind prediction of natural video quality.
\newblock \emph{{IEEE} Trans. Image Process.}, 23\penalty0 (3):\penalty0 1352--1365, 2014.

\bibitem[Saha et~al.(2023)Saha, Mishra, and Bovik]{REIQA}
Avinab Saha, Sandeep Mishra, and Alan~C. Bovik.
\newblock Re-iqa: Unsupervised learning for image quality assessment in the wild.
\newblock In \emph{{IEEE/CVF} Conference on Computer Vision and Pattern Recognition, {CVPR} 2023, Vancouver, BC, Canada, June 17-24, 2023}, pages 5846--5855. {IEEE}, 2023.

\bibitem[Sain et~al.(2023)Sain, Bhunia, Chowdhury, Koley, Xiang, and Song]{CLIPsemantic7}
Aneeshan Sain, Ayan~Kumar Bhunia, Pinaki~Nath Chowdhury, Subhadeep Koley, Tao Xiang, and Yi{-}Zhe Song.
\newblock {CLIP} for all things zero-shot sketch-based image retrieval, fine-grained or not.
\newblock In \emph{{IEEE/CVF} Conference on Computer Vision and Pattern Recognition, {CVPR} 2023, Vancouver, BC, Canada, June 17-24, 2023}, pages 2765--2775. {IEEE}, 2023.

\bibitem[Sinno and Bovik(2019)]{LIVE-VQC}
Zeina Sinno and Alan~Conrad Bovik.
\newblock Large-scale study of perceptual video quality.
\newblock \emph{{IEEE} Trans. Image Process.}, 28\penalty0 (2):\penalty0 612--627, 2019.

\bibitem[Sun et~al.(2022{\natexlab{a}})Sun, Yu, Xu, Zhou, and Chen]{GraphIQA}
Simeng Sun, Tao Yu, Jiahua Xu, Wei Zhou, and Zhibo Chen.
\newblock Graphiqa: Learning distortion graph representations for blind image quality assessment.
\newblock \emph{IEEE Transactions on Multimedia}, 2022{\natexlab{a}}.

\bibitem[Sun et~al.(2022{\natexlab{b}})Sun, Min, Lu, and Zhai]{simpleVQA}
Wei Sun, Xiongkuo Min, Wei Lu, and Guangtao Zhai.
\newblock A deep learning based no-reference quality assessment model for {UGC} videos.
\newblock In \emph{{ACM} Multimedia}, pages 856--865. {ACM}, 2022{\natexlab{b}}.

\bibitem[Tu et~al.(2021{\natexlab{a}})Tu, Wang, Birkbeck, Adsumilli, and Bovik]{UGC-VQA}
Zhengzhong Tu, Yilin Wang, Neil Birkbeck, Balu Adsumilli, and Alan~C. Bovik.
\newblock {UGC-VQA:} benchmarking blind video quality assessment for user generated content.
\newblock \emph{{IEEE} Trans. Image Process.}, 30:\penalty0 4449--4464, 2021{\natexlab{a}}.

\bibitem[Tu et~al.(2021{\natexlab{b}})Tu, Yu, Wang, Birkbeck, Adsumilli, and Bovik]{RAPIQUE}
Zhengzhong Tu, Xiangxu Yu, Yilin Wang, Neil Birkbeck, Balu Adsumilli, and Alan~C Bovik.
\newblock Rapique: Rapid and accurate video quality prediction of user generated content.
\newblock \emph{IEEE Open Journal of Signal Processing}, 2:\penalty0 425--440, 2021{\natexlab{b}}.

\bibitem[Wang et~al.(2022)Wang, Yang, Li, Liu, Wu, and Jiang]{stts}
Junke Wang, Xitong Yang, Hengduo Li, Li Liu, Zuxuan Wu, and Yu{-}Gang Jiang.
\newblock Efficient video transformers with spatial-temporal token selection.
\newblock In \emph{Computer Vision - {ECCV} 2022 - 17th European Conference, Tel Aviv, Israel, October 23-27, 2022, Proceedings, Part {XXXV}}, pages 69--86. Springer, 2022.

\bibitem[Wang et~al.(2023)Wang, Chan, and Loy]{CLIPIQA}
Jianyi Wang, Kelvin~CK Chan, and Chen~Change Loy.
\newblock Exploring clip for assessing the look and feel of images.
\newblock In \emph{Proceedings of the AAAI Conference on Artificial Intelligence}, pages 2555--2563, 2023.

\bibitem[Wang et~al.(2019)Wang, Inguva, and Adsumilli]{youtubeUGC}
Yilin Wang, Sasi Inguva, and Balu Adsumilli.
\newblock Youtube {UGC} dataset for video compression research.
\newblock In \emph{{MMSP}}, pages 1--5. {IEEE}, 2019.

\bibitem[Wu et~al.(2022{\natexlab{a}})Wu, Chen, Hou, Liao, Wang, Sun, Yan, and Lin]{FastVQA}
Haoning Wu, Chaofeng Chen, Jingwen Hou, Liang Liao, Annan Wang, Wenxiu Sun, Qiong Yan, and Weisi Lin.
\newblock {FAST-VQA:} efficient end-to-end video quality assessment with fragment sampling.
\newblock In \emph{{ECCV} {(6)}}, pages 538--554. Springer, 2022{\natexlab{a}}.

\bibitem[Wu et~al.(2022{\natexlab{b}})Wu, Liao, Chen, Hou, Wang, Sun, Yan, and Lin]{Dover}
Haoning Wu, Liang Liao, Chaofeng Chen, Jingwen Hou, Annan Wang, Wenxiu Sun, Qiong Yan, and Weisi Lin.
\newblock Disentangling aesthetic and technical effects for video quality assessment of user generated content.
\newblock \emph{CoRR}, abs/2211.04894, 2022{\natexlab{b}}.

\bibitem[Wu et~al.(2023{\natexlab{a}})Wu, Chen, Liao, Hou, Sun, Yan, Gu, and Lin]{pamifastvqa}
Haoning Wu, Chaofeng Chen, Liang Liao, Jingwen Hou, Wenxiu Sun, Qiong Yan, Jinwei Gu, and Weisi Lin.
\newblock Neighbourhood representative sampling for efficient end-to-end video quality assessment.
\newblock \emph{{IEEE} Trans. Pattern Anal. Mach. Intell.}, 45\penalty0 (12):\penalty0 15185--15202, 2023{\natexlab{a}}.

\bibitem[Wu et~al.(2023{\natexlab{b}})Wu, Chen, Liao, Hou, Sun, Yan, and Lin]{discovqa}
Haoning Wu, Chaofeng Chen, Liang Liao, Jingwen Hou, Wenxiu Sun, Qiong Yan, and Weisi Lin.
\newblock Discovqa: Temporal distortion-content transformers for video quality assessment.
\newblock \emph{{IEEE} Trans. Circuits Syst. Video Technol.}, 33\penalty0 (9):\penalty0 4840--4854, 2023{\natexlab{b}}.

\bibitem[Wu et~al.(2023{\natexlab{c}})Wu, Zhang, Liao, Chen, Hou, Wang, Sun, Yan, and Lin]{maxwell}
Haoning Wu, Erli Zhang, Liang Liao, Chaofeng Chen, Jingwen Hou, Annan Wang, Wenxiu Sun, Qiong Yan, and Weisi Lin.
\newblock Towards explainable in-the-wild video quality assessment: {A} database and a language-prompted approach.
\newblock In \emph{Proceedings of the 31st {ACM} International Conference on Multimedia, {MM} 2023, Ottawa, ON, Canada, 29 October 2023- 3 November 2023}, pages 1045--1054. {ACM}, 2023{\natexlab{c}}.

\bibitem[Wu et~al.(2023{\natexlab{d}})Wu, Hu, Xiao, Deng, Li, Chen, and Li]{NTIRE1}
Wei Wu, Shuming Hu, Pengxiang Xiao, Sibin Deng, Yilin Li, Ying Chen, and Kai Li.
\newblock Video quality assessment based on swin transformer with spatio-temporal feature fusion and data augmentation.
\newblock In \emph{{IEEE/CVF} Conference on Computer Vision and Pattern Recognition, {CVPR} 2023 - Workshops, Vancouver, BC, Canada, June 17-24, 2023}, pages 1846--1854. {IEEE}, 2023{\natexlab{d}}.

\bibitem[Xing et~al.(2022)Xing, Wang, Wang, Li, and Zhu]{starVQA}
Fengchuang Xing, Yuan{-}Gen Wang, Hanpin Wang, Leida Li, and Guopu Zhu.
\newblock Starvqa: Space-time attention for video quality assessment.
\newblock In \emph{2022 {IEEE} International Conference on Image Processing, {ICIP} 2022, Bordeaux, France, 16-19 October 2022}, pages 2326--2330. {IEEE}, 2022.

\bibitem[Xiong et~al.(2023)Xiong, Pavez, Ortega, and Adsumilli]{transcode}
Xin Xiong, Eduardo Pavez, Antonio Ortega, and Balu Adsumilli.
\newblock Rate-distortion optimization with alternative references for {UGC} video compression.
\newblock In \emph{{IEEE} International Conference on Acoustics, Speech and Signal Processing {ICASSP} 2023, Rhodes Island, Greece, June 4-10, 2023}, pages 1--5. {IEEE}, 2023.

\bibitem[Xu et~al.(2021)Xu, Li, Zhou, Zhou, Wang, and Chen]{Internet}
Jiahua Xu, Jing Li, Xingguang Zhou, Wei Zhou, Baichao Wang, and Zhibo Chen.
\newblock Perceptual quality assessment of internet videos.
\newblock In \emph{{MM} '21: {ACM} Multimedia Conference, Virtual Event, China, October 20 - 24, 2021}, pages 1248--1257. {ACM}, 2021.

\bibitem[Xu et~al.(2023)Xu, Zhang, Wei, Hu, and Bai]{CLIPsemantic4}
Mengde Xu, Zheng Zhang, Fangyun Wei, Han Hu, and Xiang Bai.
\newblock Side adapter network for open-vocabulary semantic segmentation.
\newblock In \emph{{CVPR}}, pages 2945--2954. {IEEE}, 2023.

\bibitem[Xue et~al.(2019)Xue, Chen, Wu, Wei, and Freeman]{enhancement2}
Tianfan Xue, Baian Chen, Jiajun Wu, Donglai Wei, and William~T. Freeman.
\newblock Video enhancement with task-oriented flow.
\newblock \emph{Int. J. Comput. Vis.}, 127\penalty0 (8):\penalty0 1106--1125, 2019.

\bibitem[Yan et~al.(2023)Yan, Dong, Zhang, and Tang]{CLIPsemantic6}
Shuanglin Yan, Neng Dong, Liyan Zhang, and Jinhui Tang.
\newblock Clip-driven fine-grained text-image person re-identification.
\newblock \emph{{IEEE} Trans. Image Process.}, 32:\penalty0 6032--6046, 2023.

\bibitem[Yang et~al.(2020)Yang, Mentzer, Gool, and Timofte]{compression}
Ren Yang, Fabian Mentzer, Luc~Van Gool, and Radu Timofte.
\newblock Learning for video compression with hierarchical quality and recurrent enhancement.
\newblock In \emph{2020 {IEEE/CVF} Conference on Computer Vision and Pattern Recognition, {CVPR} 2020, Seattle, WA, USA, June 13-19, 2020}, pages 6627--6636. Computer Vision Foundation / {IEEE}, 2020.

\bibitem[Yang et~al.(2021)Yang, Li, and Liu]{TTLIQA}
Xiaohan Yang, Fan Li, and Hantao Liu.
\newblock {TTL-IQA:} transitive transfer learning based no-reference image quality assessment.
\newblock \emph{{IEEE} Trans. Multim.}, 23:\penalty0 4326--4340, 2021.

\bibitem[Ying et~al.(2021{\natexlab{a}})Ying, Mandal, Ghadiyaram, and Bovik]{PVQ}
Zhenqiang Ying, Maniratnam Mandal, Deepti Ghadiyaram, and Alan Bovik.
\newblock Patch-vq:'patching up'the video quality problem.
\newblock In \emph{Proceedings of the IEEE/CVF conference on computer vision and pattern recognition}, pages 14019--14029, 2021{\natexlab{a}}.

\bibitem[Ying et~al.(2021{\natexlab{b}})Ying, Mandal, Ghadiyaram, and Bovik]{LSVQ}
Zhenqiang Ying, Maniratnam Mandal, Deepti Ghadiyaram, and Alan~C. Bovik.
\newblock Patch-vq: 'patching up' the video quality problem.
\newblock In \emph{{CVPR}}, pages 14019--14029. Computer Vision Foundation / {IEEE}, 2021{\natexlab{b}}.

\bibitem[You(2021)]{LSCT}
Junyong You.
\newblock Long short-term convolutional transformer for no-reference video quality assessment.
\newblock In \emph{{MM} '21: {ACM} Multimedia Conference, Virtual Event, China, October 20 - 24, 2021}, pages 2112--2120. {ACM}, 2021.

\bibitem[Yuan et~al.(2023)Yuan, Kong, Zheng, Sun, and Wen]{VQT}
Kun Yuan, Zishang Kong, Chuanchuan Zheng, Ming Sun, and Xing Wen.
\newblock Capturing co-existing distortions in user-generated content for no-reference video quality assessment.
\newblock In \emph{{ACM} Multimedia}, pages 1098--1107. {ACM}, 2023.

\bibitem[Zhang et~al.(2022{\natexlab{a}})Zhang, Wang, Tang, Li, and Kwong]{HVSvist}
Ao{-}Xiang Zhang, Yuan{-}Gen Wang, Weixuan Tang, Leida Li, and Sam Kwong.
\newblock {HVS} revisited: {A} comprehensive video quality assessment framework.
\newblock \emph{CoRR}, abs/2210.04158, 2022{\natexlab{a}}.

\bibitem[Zhang et~al.(2023{\natexlab{a}})Zhang, Ran, Tang, and Wang]{robustVQA}
Ao{-}Xiang Zhang, Yu Ran, Weixuan Tang, and Yuan{-}Gen Wang.
\newblock Vulnerabilities in video quality assessment models: The challenge of adversarial attacks.
\newblock \emph{CoRR}, abs/2309.13609, 2023{\natexlab{a}}.

\bibitem[Zhang et~al.(2022{\natexlab{b}})Zhang, Guo, Zhang, Li, Miao, Cui, Qiao, Gao, and Li]{CLIPsemantic5}
Renrui Zhang, Ziyu Guo, Wei Zhang, Kunchang Li, Xupeng Miao, Bin Cui, Yu Qiao, Peng Gao, and Hongsheng Li.
\newblock Pointclip: Point cloud understanding by {CLIP}.
\newblock In \emph{{IEEE/CVF} Conference on Computer Vision and Pattern Recognition, {CVPR} 2022, New Orleans, LA, USA, June 18-24, 2022}, pages 8542--8552. {IEEE}, 2022{\natexlab{b}}.

\bibitem[Zhang et~al.(2023{\natexlab{b}})Zhang, Zhai, Wei, Yang, and Ma]{LIQE}
Weixia Zhang, Guangtao Zhai, Ying Wei, Xiaokang Yang, and Kede Ma.
\newblock Blind image quality assessment via vision-language correspondence: A multitask learning perspective.
\newblock In \emph{Proceedings of the IEEE/CVF Conference on Computer Vision and Pattern Recognition}, pages 14071--14081, 2023{\natexlab{b}}.

\bibitem[Zhang et~al.(2023{\natexlab{c}})Zhang, Wu, Sun, Tu, Lu, Min, Chen, and Zhai]{MDVQA}
Zicheng Zhang, Wei Wu, Wei Sun, Danyang Tu, Wei Lu, Xiongkuo Min, Ying Chen, and Guangtao Zhai.
\newblock {MD-VQA:} multi-dimensional quality assessment for {UGC} live videos.
\newblock In \emph{{CVPR}}, pages 1746--1755. {IEEE}, 2023{\natexlab{c}}.

\bibitem[Zhao et~al.(2023{\natexlab{a}})Zhao, Yuan, Sun, Li, and Wen]{QPT}
Kai Zhao, Kun Yuan, Ming Sun, Mading Li, and Xing Wen.
\newblock Quality-aware pre-trained models for blind image quality assessment.
\newblock \emph{CoRR}, abs/2303.00521, 2023{\natexlab{a}}.

\bibitem[Zhao et~al.(2023{\natexlab{b}})Zhao, Yuan, Sun, and Wen]{zoomVQA}
Kai Zhao, Kun Yuan, Ming Sun, and Xing Wen.
\newblock Zoom-vqa: Patches, frames and clips integration for video quality assessment.
\newblock In \emph{{IEEE/CVF} Conference on Computer Vision and Pattern Recognition, {CVPR} 2023 - Workshops, Vancouver, BC, Canada, June 17-24, 2023}, pages 1302--1310. {IEEE}, 2023{\natexlab{b}}.

\bibitem[Zheng et~al.(2022)Zheng, Tu, Zeng, Bovik, and Fan]{VIQE}
Qi Zheng, Zhengzhong Tu, Xiaoyang Zeng, Alan~C Bovik, and Yibo Fan.
\newblock A completely blind video quality evaluator.
\newblock \emph{IEEE Signal Processing Letters}, 29:\penalty0 2228--2232, 2022.

\bibitem[Zhou et~al.(2023)Zhou, Zhou, and Qiu]{zhou2023blind}
Zehong Zhou, Fei Zhou, and Guoping Qiu.
\newblock Blind image quality assessment based on separate representations and adaptive interaction of content and distortion.
\newblock \emph{IEEE Transactions on Circuits and Systems for Video Technology}, 2023.

\bibitem[Zhu et~al.(2023)Zhu, Jin, Yang, Wu, and Wang]{CLIPsemantic3}
Jun Zhu, Jiandong Jin, Zihan Yang, Xiaohao Wu, and Xiao Wang.
\newblock Learning {CLIP} guided visual-text fusion transformer for video-based pedestrian attribute recognition.
\newblock In \emph{{CVPR} Workshops}, pages 2626--2629. {IEEE}, 2023.

\end{thebibliography}
